\documentclass[12pt]{article}

\font\grande=cmr9.5 scaled \magstep4
\font\medio=cmr9.5 scaled \magstep2
\outer\def\beginsection#1\par{\medbreak\bigskip
      \message{#1}\leftline{\bf#1}\nobreak\medskip
\vskip-\parskip
      \noindent}
\usepackage{graphicx} 
\begin{document}
\bibliographystyle {unsrt}

\titlepage

\begin{flushright}
CERN-PH-TH/2010-150
\end{flushright}

\vspace{15mm}
\begin{center}
{\grande Magnetic field contribution}\\
\vspace{5mm}
{\grande to the last electron-photon scattering}\\
\vspace{1.5cm}
 Massimo Giovannini 
 \footnote{Electronic address: massimo.giovannini@cern.ch} \\
\vspace{1cm}
{{\sl Department of Physics, 
Theory Division, CERN, 1211 Geneva 23, Switzerland }}\\
\vspace{0.5cm}
{{\sl INFN, Section of Milan-Bicocca, 20126 Milan, Italy}}
\vspace*{2cm}

\end{center}

\vskip 1cm
\centerline{\medio  Abstract}
When the cosmic microwave photons scatter electrons just prior to the decoupling of matter and radiation, magnetic fields do contribute to the Stokes matrix as well as to the scalar, vector and tensor components of the transport equations for the brightness perturbations.  The magnetized electron-photon scattering  is hereby discussed  in general terms by including, for the first time, the contribution of magnetic fields with arbitrary direction and in the  presence of the scalar, vector and tensor modes of the geometry. The propagation of relic vectors and relic gravitons is discussed for a varying magnetic field orientation and for different photon directions.  The source terms of the transport equations in the presence of the relativistic fluctuations of the geometry are also explicitly averaged over the magnetic field orientations and the problem of a consistent account of the small-scale and large-scale magnetic field is briefly outlined.
\noindent

\vspace{5mm}
\vfill
\newpage
\renewcommand{\theequation}{1.\arabic{equation}}
\setcounter{equation}{0}
\section{Formulation of the problem}
\label{sec1}
The last electron-photon scattering is customarily discussed without the additional complication of a magnetic field.  In numerical codes as well as 
in analytical estimates, the collisional contributions are evaluated as if electrons and ions were free right before last scattering\cite{chandra}.  The relativistic fluctuations of the geometry are included in the classic transport problem \cite{chandra} either by using specific gauges \cite{mab} or with fully gauge-invariant 
methods.  
The resulting equations including both the source terms coming from 
electron-photon scattering and the relativistic fluctuations of the geometry 
form the set of transport equations which can be solved within various approaches either by truncating the system at a specific (maximal) multipole \cite{mab} or by using the integration 
along the line of sight \cite{sel1}. One of the consequences of the consistent solution of the system of transport equations are estimates of the temperature and polarization inhomogeneities 
of the Cosmic Microwave Background (CMB in what follows).
The recent WMAP 7 data \cite{WMAP7a,WMAP7b,WMAP7c,WMAP7d,WMAP7e,WMAP7f} 
are able to constrain the vanilla $\Lambda$CDM scenario (where $\Lambda$ stands for the dark-energy component
and CDM for the cold dark matter component). In the near future the $\Lambda$CDM scenario\footnote{$\Lambda$ stands for the 
dark-energy component while CDM denotes the cold dark matter component.} will be tested not only in its minimal version but in its non-minimal extensions ranging from the addition of a stochastic background of relic tensor modes of the geometry to large-scale magnetic fields \cite{mg1a,mg2a}.

In recent years there has been mounting evidence 
of the role played by magnetic fields at large scales \cite{mg1a,mg2a}. Why should magnetic fields be assumed in various processes ranging from star formation to cluster dynamics and completely neglected prior to last scattering? Why are magnetic fields overlooked in CMB physics while they are  observed in galaxies clusters, superclusters and high-redshift quasars? 
There are no reasons for doing so unless one would implicitly assume that large-scale magnetism suddenly arose between hydrogen recombination and, say, the gravitational 
collapse of the protogalaxy. While it might well be that the latter situation is the one preferred by nature, it would be nice to have some direct empirical evidence less biased by speculations.  
To comply with the latter program, a specific approach has been tailored through the last few years \cite{mg2a} (see also \cite{mg1}). The idea is, in a nutshell, to introduce consistently 
large-scale magnetic fields in all the steps leading to the estimate of CMB anisotropies and polarization.  So far the program undertaken in \cite{mg2a} led to various results
\begin{itemize}
\item{} the large-scale magnetic fields have been included both at the level of the initial conditions as well as the level of the evolution equations for the standard adiabatic mode
 and for the other entropic initial conditions \cite{mg1};
\item{} the temperature and polarization anisotropies 
induced by the magnetized (adiabatic and entropic) initial conditions have been computed \cite{mg2};
\item{} the parameters of the magnetized background have been 
estimated (for the first time) in \cite{mg3} by using  the 
TT and TE correlations\footnote{Following the standard shorthand 
terminology the TT correlations denote the temperature 
autocorrelations while the TE correlations denote the cross-correlation between the temperature and the E-mode polarization.} measured by the WMAP collaboration.
\end{itemize}
There exist other approaches to the interplay between large-scale magnetic fields and CMB anisotropies (see \cite{vt1,vt2,vt3,vt4,vt5} for an incomplete list of references; see \cite{mg2a} for a more thorough account of earlier results). The common characteristic of those approaches has been to 
neglect the scalar modes of the geometry and to focus the attention to the tensor and vector modes.  The recent results 
\cite{mg1,mg2,mg3} show, in contrast with previous guesses, that large scale magnetic fields alter the initial conditions 
and the dynamics of the scalar modes of the geometry. They 
consequently distort, in a computable manner, the temperature and polarization anisotropies. 

A limitation common to nearly all studies on pre-decoupling magnetism has been so far the total absence of the effect of the magnetic field in the process of electron-photon scattering. In 
\cite{mg1,mg2,mg3}, for instance, the magnetic fields are 
included in the initial conditions and in all the relevant governing equations.  The electron-photon scattering, however, is assumed to take place as if the magnetic fields were absent. The potential smallness of the effects does not justify its neglect since 
diverse small effects are often claimed to be detectable because of the purported control we now have on CMB foregrounds 
\cite{NG}.  

The consistent inclusion of magnetic fields 
in electron-photon scattering modifies qualitatively the standard lore since the geodesics of electrons and ions are be modified by the presence of the Lorentz force term in curved backgrounds. Absent the contribution of the magnetic 
field, the motion of electrons and ions depends only upon the 
incident electric field; but when the magnetic field is 
included the classic treatment (see, for instance \cite{chandra}) must be adapted to the new situation. 

The neglect of the role of the magnetic field in the electron-photon scattering has been recently relaxed  in the guiding centre approximation \cite{mgpol1,mgpol2,mgpol3,mgpol4}, and for a specified magnetic field orientation. The argument for keeping the direction fixed was essentially practical and in the present paper a general treatment will by developed along a twofold perspective 
\begin{itemize}
\item{} the orientation of the magnetic field will be kept arbitrary so that the matrix
elements either in in the Jones or in the Mueller calculus will depend not only 
upon the directions of the incident and of the outgoing radiation but also 
on the magnetic field orientation;
\item{} after including the magnetic field in the scattering process the 
source terms for the transport equations of the scalar, vector and tensor 
modes of the geometry will be deduced explicitly.
\end{itemize}
The latter analysis is still lacking both in the present and in the 
earlier literature.  The magnetized electron-photon scattering is often 
required in diverse astrophyiscal situations like in the 
physics of magnetized sun spots \cite{sun}, or  
 the theory of synchrotron emission \cite{sync,sync2}
whose  results 
cannot be directly used since prior at last scattering  
electrons and ions are notoriously non-relativistic. 
Conversely some studies 
involving directly Thomson scattering in a 
magnetized environment \cite{TS} do not incorporate 
the fluctuations of the geometry and are also obtained 
using a preferential magnetic field orientation. 

The layout of this paper is therefore the following. 
In section \ref{sec2}  the tenets of the Mueller and Jones 
calculus will be reviewed and the matrix elements for the 
magnetized electron-photon scattering presented.  Section \ref{sec3} introduces  the scalar, vector and tensor components 
of the brightness perturbations and the calculation of the collisionless part of the transport equations.
The full scalar, vector and tensor transport equations will be discussed, respectively, in sections \ref{sec4}, 
\ref{sec5} and \ref{sec6}. Section \ref{sec7}  contains the concluding remarks. Explicit expressions involving  all the  relevant matrix elements both in the Jones and in the Mueller approaches have been collected in the appendices \ref{APPA} and \ref{APPB}. 

\renewcommand{\theequation}{2.\arabic{equation}}
\setcounter{equation}{0}
\section{Mueller and Jones calculus}
\label{sec2}
\begin{figure}[!ht]
\centering
\includegraphics[height=7cm]{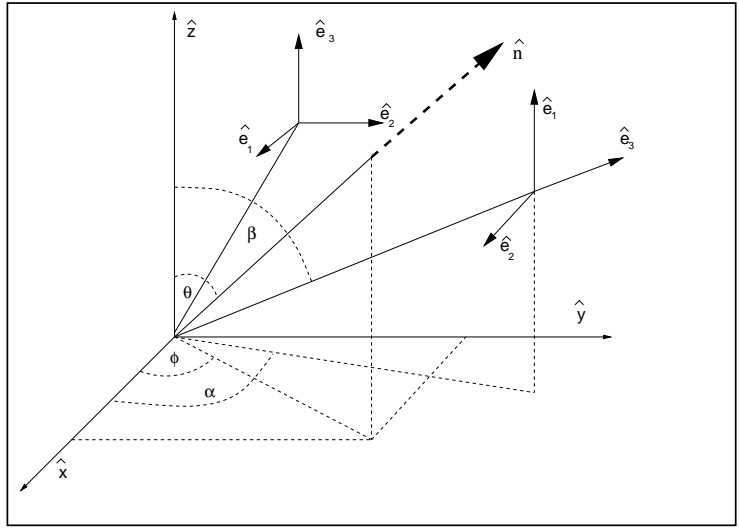}
\caption[a]{Schematic view of the relation between the coordinate system defining the scattered 
radiation field and the local frame of reference defining the direction of the magnetic field.}
\label{F1}      
\end{figure}

In the Mueller calculus the Stokes parameters are organized in a four-dimensional (Mueller) column vector whose components 
are exactly the four Stokes parameters, i.e. $I$, $Q$, $U$ and $V$. In the Jones calculus the electric fields of the 
wave are organized in a two-dimensional column vector and the Stokes parameters are effectively derived 
quantities (see \cite{robson} for an introduction to the Mueller and Jones approaches). Hereunder a hybrid approach shall be employed. The polarization tensor  
${\mathcal P}_{ij}= {\mathcal P}_{ji}= E_{i}\, E_{j}^{*}$ can be organized in a Stokes matrix whose explicit form is:
\begin{equation}
{\mathcal P}= \frac{1}{2} \left(\matrix{ I + Q
& U - i V &\cr
U + i V & I - Q &\cr}\right) = \frac{1}{2} \left( I\,{\bf 1} +U\, \sigma_{1} + V\, \sigma_2 +Q\, \sigma_3 \right),
\label{PM}
\end{equation}
where ${\bf 1}$ denotes the identity matrix while $\sigma_{1}$, $\sigma_{2}$ and $\sigma_{3}$ are the three 
Pauli matrices.  Sometimes the Stokes matrix ${\mathcal P}$ is separated in a traceless  part (i.e. the polarization matrix) supplemented by the identity matrix multiplying the intensity of the radiation field: this separation shall not be employed here. 
The orientation of the coordinate system is illustrated in 
Fig. \ref{F1}. The radial, azimuthal and polar 
directions are 
\begin{eqnarray}
&& \hat{r} = (\cos{\varphi} \sin{\vartheta},\, \sin{\varphi} \sin{\vartheta},\, \cos{\vartheta}),
\nonumber\\
&& \hat{\vartheta} = (\cos{\varphi} \cos{\vartheta},\, \sin{\varphi} \cos{\vartheta},\, -\sin{\vartheta}),
\nonumber\\
&& \hat{\varphi} = ( - \sin{\varphi},\, \cos{\varphi},\, 0),
\label{CS1}
\end{eqnarray}
implying that $\hat{r} \times \hat{\vartheta} = \hat{\varphi}$. 
Photons propagate radially and $\hat{n}=(\vartheta, \varphi)$ denotes the direction of the scattered photon 
while $\hat{n}' = (\vartheta', \varphi')$ is the direction of the incoming photon; similarly $\mu = \cos{\vartheta}$ and 
$\nu = \cos{\vartheta'}$.  

When the photons impinge the electrons in a magnetized 
environment the magnetic field can be treated in the guiding centre approximation. Denoting with $\vec{B}$ the comoving magnetic field intensity the guiding centre approximation \cite{alfven,boyd} stipulates
\begin{equation}
B_{i}(\vec{x},\tau) \simeq B_{i}(\vec{x}_{0}, \tau)  + (x^{j} - x_{0}^{j}) \partial_{j} B_{i} +...
\label{CS2}
\end{equation}
where the ellipses stand for the higher orders in the gradients leading, both, to curvature and drift corrections which will be neglected in this investigation. The scales one must therefore compare are $|\vec{x}_{0}| = L_{0}$, $|\vec{x} - \vec{x}_{0}| = L$,
$\lambda^{(\mathrm{rec})}_{\gamma}$ (the wavelength of the incident radiation at the recombination epoch) 
and $H^{-1}_{\mathrm{rec}}$ (i.e. the Hubble rate at recombination). It is easy to appreciate that $\lambda^{(\mathrm{rec})}_{\gamma}= {\mathcal O}(\mu\mathrm{m})$ implying that
\begin{equation}
H^{-1}_{\mathrm{rec}} \simeq L \gg L_{0} \gg \lambda_{\gamma}^{(\mathrm{rec})}.
\label{CS3}
\end{equation}
Equation (\ref{CS3}) implies that, for the purposes of the contribution of the magnetic field to the Stokes matrix the spatial gradients can be neglected while they cannot be neglected when estimating the effects of the large-scale inhomogeneities 
of the magnetic field. In spite of the fact that the contribution of the 
spatial gradients can be neglected in the first approximation, still the direction of the magnetic field should be appropriately taken into account. Consequently it is necessary to introduce a local basis which will define for us the magnetic field direction:
\begin{eqnarray}
&&\hat{e}_{1} = (\cos{\alpha} \cos{\beta},\, \sin{\alpha} \cos{\beta}, \, - \sin{\beta}),
\nonumber\\
&&\hat{e}_{2}= (- \sin{\alpha},\, \cos{\alpha},\, 0),
\nonumber\\
&&\hat{e}_{3} = (\cos{\alpha} \sin{\beta}, \sin{\alpha} \sin{\beta}, \cos{\beta}).
\label{CS4}
\end{eqnarray}
The basis of Eq. (\ref{CS4}) {\em local} since 
it accounts for the direction of the magnetic field over the typical scales involved in the electron-photon scattering. 
Once the direction of the local magnetic field has been fixed, the motion of the electrons and of the ions will follow the appropriate geodesics holding for charged particles in a gravitational field. 

For the calculation of the scattering matrix the magnetic field can be aligned along $\hat{e}_{3}$. The latter choice is purely conventional and it does not prevent from varying 
arbitrarily the direction of the magnetic field with respect either to the direction of propagation of the photons 
or to the direction of propagation of the other fluctuations of the geometry. Consider, as an example,  a situation which will be treated later on in greater detail, i.e. the case where 
a relic vector mode of the geometry \footnote{The same discussion, 
with the due differences,  can be repeated in the case of the scalar or tensor modes of the geometry. Here the case of the vector 
modes is just selected for sake of illustration.}
 propagates along the direction $\hat{k}$.  Since $\hat{k}$ cooincides 
also, by definition, with the direction of the Fourier wavevector the whole problem will be characterized by 
\begin{itemize}
\item{} $(\hat{n}\cdot \hat{k})$, i.e. the projection of the photon momentum along the direction of propagation of the relic vector;
\item{} $(\hat{e}_{3}\cdot \hat{k})$, i.e. the projection of the magnetic field direction along the direction of propagation 
of the relic vector.
\end{itemize}
In the mentioned example we can choose, without loss of generality, $\hat{k} = \hat{z}$ and the two physical polarizations 
of the relic vector will then be defined in the $\hat{x}-\hat{y}$ plane. In this situation 
$\cos{\vartheta} = \hat{k}\cdot \hat{n}$ and $\cos{\alpha} = \hat{k} \cdot\hat{e}_{3}$.
The direction $\hat{e}_{3}$ does not coincide, in general, with $\hat{z}$.
For instance if $\alpha = \beta = - \pi/2$, $\hat{e}_{3}$ coincides with $\hat{e}_{y}$ while for $\alpha=0$ and $\beta = \pi/2$ 
$\hat{e}_{3}$ coincides with $\hat{e}_{x}$.
This simple example shows explicitly that since the direction of $\hat{e}_{3}$ is arbitrary,  the 
orientation of the magnetic field is also generic. Such an arbitrariness entails the dependence of the 
scattering matrix upon two supplementary angles. In total the 
Stokes matrix will then depend 
upon the {\em two angles} defining the direction of the scattered radiation, 
the {\em two angles} defining the direction of the incident radiation and the {\em two angles} defining the direction 
of the magnetic field. The Stokes matrix will then depend overall upon six angles: $(\vartheta,\varphi)$ (for the directions of the scattered photons), $(\vartheta',\varphi')$ (for the directions of the incident photons) and $(\alpha,\beta)$ for the magnetic field direction. 
The schematic relation between the direction of the scattered radiation and the local frame defined by Eq. 
(\ref{CS4}) is summarized in Fig. \ref{F1}.  The (thick) dashed line denotes the direction of $\hat{n}$, i.e. 
the direction of propagation of the radiation field. In Fig. \ref{F1} the two different alignments of $\hat{e}_{3}$ are just meant 
to illustrate the effective arbitrariness of the magnetic field orientation. 

In the dipole approximation the scattered electric field can be computed as the composition 
of the scattered electric fields due to the electrons and to the ions:
\begin{equation} 
\vec{E}^{\mathrm{out}}_{(\mathrm{e})} = - e \frac{\vec{r} \times [\vec{r} \times \vec{a}_{(\mathrm{e})}]}{r^3},\qquad 
\vec{E}^{\mathrm{out}}_{(\mathrm{i})} =  e \frac{\vec{r} \times[\vec{r} \times \vec{a}_{(\mathrm{i})}]}{r^3},
\label{sc1}
\end{equation}
where $\vec{a}_{(\mathrm{e})}$ and $\vec{a}_{(\mathrm{i})}$ are, respectively, the accelerations for the electrons 
and for the ions. In the local frame defined by Eq. (\ref{CS4}) the vector $\vec{A} = (\vec{a}_{(\mathrm{e})}- 
\vec{a}_{(\mathrm{i})})$ the vector can be decomposed as 
 $\vec{A} = (A_{1} \hat{e}_{1} + A_{2} \hat{e}_{2} + A_{3} \hat{e}_{3})$. Denoting with  $E_{1}= (\vec{E}\cdot\hat{e}_{1})$, $E_{2} = (\vec{E} \cdot\hat{e}_{2})$ and $E_{3} =(\vec{E} \cdot\hat{e}_{2})$ the components of the electric fields of the incident 
 radiation in the local frame we have that from the geodesics of electrons and ions 
\begin{eqnarray}
&& A_{1} =  
\frac{\omega_{\mathrm{pe}}^2}{4 \pi  n_{0}} \, \zeta(\omega) \biggl[ \Lambda_{1} E_{1} - i f_{\mathrm{e}} \Lambda_{2} E_{2} \biggr],
\label{sc2}\\
&& A_{2} = \frac{\omega_{\mathrm{pe}}^2}{4 \pi  n_{0}} \, 
\zeta(\omega) \biggl[ \Lambda_{1} E_{2} + i f_{\mathrm{e}} \Lambda_{2} E_{1} \biggr],
\label{sc3}\\
&& A_{3} =  - \frac{\omega_{\mathrm{pe}}^2}{4 \pi  n_{0}} \Lambda_{3} E_{3},
\label{sc4}
\end{eqnarray}
where because of the global neutrality of the plasma, $n_{0} = \tilde{n}_{0} a^3$ is the common comoving concentration 
of electrons and ions;  $\omega_{\mathrm{Be,\,i}}$ and $\omega_{\mathrm{pe,\,i}}$ 
denote respectively the Larmor and plasma frequencies 
for electrons (and ions) 
\begin{equation}
\omega_{\mathrm{Be,\,i}} = \frac{e \vec{B}\cdot\hat{e}_{3}}{m_{\mathrm{e,\,i}} a}, \qquad 
\omega_{\mathrm{pe,\,i}} =\sqrt{\frac{4 \pi e^2 n_{0}}{ m_{\mathrm{e,\,i}} a}},
\label{FREQ}
\end{equation}
where $m_{\mathrm{e,\,i}}$ denote either the electron 
or the ion mass depending upon the relative subscript and $a(\tau)$ is the scale factor of a conformally flat geometry of Friedmann-Robertson-Walker type 
whose line element and metric tensor are defined as
\begin{equation}
ds^2 = g_{\mu\nu} dx^{\mu} dx^{\nu} = a^2(\tau)[ d\tau^2 - d\vec{x}^2],\qquad g_{\mu\nu} = a^2(\tau) \eta_{\mu\nu}.
\label{metric}
\end{equation}
In Eqs. (\ref{sc2}), (\ref{sc3}) and (\ref{sc4}) the functions $\Lambda_{i}$ (with $i = 1, 2, 3$) as well as $\zeta(\omega)$ and  all depend upon the angular frequency of the photon (i.e. $\omega = 2 \pi \nu$) and are defined as:
\begin{eqnarray}
&&\Lambda_{1}(\omega) = 1 + 
\biggl(\frac{\omega^2_{\mathrm{pi}}}{\omega^2_{\mathrm{pe}}}\biggr) \biggl( 
\frac{ \omega^2 - \omega^2_{\mathrm{Be}}}{\omega^2 - \omega^2_{\mathrm{Bi}}}\biggr),
\nonumber\\
&& \Lambda_{2}(\omega) = 1 - \biggl(\frac{\omega^2_{\mathrm{pi}}}{\omega^2_{\mathrm{pe}}}\biggr)
\biggl(\frac{\omega_{\mathrm{Bi}}}{\omega_{\mathrm{Be}}}\biggr) \biggl( 
\frac{ \omega^2 - \omega_{\mathrm{Be}}^2}{\omega^2 - \omega^2_{\mathrm{Bi}}}\biggr),
\nonumber\\
&& \Lambda_{3}(\omega) = 1 + \biggl(\frac{\omega^2_{\mathrm{pi}}}{\omega^2_{\mathrm{pe}}}\biggr), 
\nonumber\\
&& \zeta(\omega) = \frac{\omega^2}{\omega_{\mathrm{Be}}^2 - \omega^2} = \frac{1}{f_{\mathrm{e}}^2(\omega) -1}, \qquad f_{\mathrm{e}}(\omega) =  \biggl(\frac{\omega_{\mathrm{Be}}}{\omega}\biggr).
\label{LAM3}
\end{eqnarray}
The scale factor $a(\tau)$ appears explicitly in Eqs. (\ref{FREQ}) since 
the mass of the (non relativistic) species breaks the conformal invariance of the system of equations. Indeed, Eqs. (\ref{sc2}), (\ref{sc3}) and (\ref{sc4}) 
follow from the geodesics of charged species in the conformally flat metric 
of Eq. (\ref{metric})
where, for a generic massive particle, the mass shell condition implies that $g_{\alpha\beta} P^{\alpha} P^{\beta} = m^2$ ($P^{\alpha} = m u^{\alpha}$ is the canonical momentum and $u^{\alpha}$ the four-velocity). Recalling that the comoving 
three-momentum $\vec{q}$ is defined as $\vec{q} = a \vec{p}$ where $\delta_{ij} p^{i} p^{j} = - g_{ij} P^{i} P^{j}$, the 
comoving three-velocity is given by $\vec{v} = \vec{q}/\sqrt{q^2 + m^2 a^2}$. Since the electrons are 
non-relativistic at last scattering $\vec{q} = m a \vec{v}$ and this is, ultimately, the rationale for the appearance 
of the scale factors in the explicit expressions of the Larmor and plasma frequencies for the electrons and for 
the ions\footnote{For more explicit discussions of these points see \cite{mg3,mgpol1}. Note that Eq. (A12) of 
\cite{mgpol1} contains few trivial typos which have been corrected in the archive version of the same paper.}.
The numerical value of $f_{\mathrm{e}}(\omega)$ for typical cosmological parameters is given by 
\begin{equation}
f_{\mathrm{e}}(\omega) = \biggl(\frac{\omega_{\mathrm{Be}}}{\omega}\biggr)
 = 2.79 \times 10^{-12} \biggl(\frac{B}{\mathrm{nG}}\biggr)
 \biggl(\frac{\mathrm{GHz}}{\nu}\biggr) (z_{*} +1)\ll 1,
 \label{FE}
 \end{equation}
where $z_{*}$ is the redshift to last scattering, i.e. $z_{*} = 1090.79_{-0.92}^{+0.94}$ according to the WMAP-7yr data \cite{WMAP7f}. In Eq. (\ref{FE}) $B= |\hat{e}_{3}\cdot\vec{B}|$; grossly speaking the typical values of $\nu$ and $B$ appearing 
 in Eq. (\ref{FE}) do correspond, respectively, to the (very minimal) value of the frequency channel of CMB experiments and to the maximal value of the comoving magnetic field allowed by the distortions of the temperature autocorrelations 
 and of the cross-correlations between temperature and polarization. The evolution of the Stokes matrix ${\mathcal P}$ can be formally written as\footnote{The dagger in Eq. (\ref{EV1}) defines, as usual, the complex conjugate of the transposed matrix.}
\begin{equation}
 \frac{d {\mathcal P}}{d\tau} + \epsilon' {\mathcal P} =  \frac{3 \epsilon'}{16 \pi} \int M(\Omega,\Omega',\alpha,\beta)\, {\mathcal P}(\Omega,\Omega') \,M^{\dagger}(\Omega,\Omega',\alpha,\beta),
\label{EV1}
\end{equation}
where $d\Omega' = d\cos{\vartheta'} \,d\varphi'$  and, defining 
the rate of electron-photon scattering $\Gamma_{\gamma\mathrm{e}}$,
\begin{equation}
\epsilon' =a \Gamma_{\gamma\mathrm{e}} = a \tilde{n}_{\mathrm{0}} x_{\mathrm{e}} \sigma_{\mathrm{e}\gamma}, \qquad \sigma_{\gamma\mathrm{e}} = \frac{8}{3} \pi r_{\mathrm{e}}^2, \qquad r_{\mathrm{e}}= \frac{e^2}{m_{\mathrm{e}}}
\label{diffop}
\end{equation}
is the differential optical depth. 
At the right of Eq. (\ref{EV1})
 the matrix $M(\Omega,\Omega',\alpha,\beta)$ is a $2\times2$ and the  
 four entries of the matrix $M(\Omega,\Omega',\alpha,\beta)$ are separately reported in appendix \ref{APPA}. 
 From the matrix elements 
 of $M(\Omega,\Omega',\alpha,\beta)$ it is immediately possible to derive the evolution equations for the Stokes parameters 
 in the Mueller form, namely, 
 \begin{equation}
 \frac{d {\mathcal I}}{d\tau} + \epsilon' {\mathcal I} = \frac{3 \epsilon'}{32\pi} \int d\Omega' \,\, {\mathcal T}(\Omega, \Omega',\alpha,\beta) {\mathcal I}(\Omega'),
 \label{EV2}
 \end{equation}
where ${\mathcal I}$ is a column matrix whose entries are, respectively, $I$, $Q$, $U$ and $V$. The entries of the $4\times4$ Mueller matrix will be denoted as ${\mathcal T}_{ij}(\Omega,\Omega',\alpha,\beta)$ where $i$ and $j$ run over the various Stokes parameters $I$, $Q$, $U$ and $V$ and are reported in 
appendix \ref{APPA} (see, in particular, Eqs. (\ref{EV3})--
(\ref{EV18})). Finally, in terms of the matrix the evolution equations
of the different Stokes parameters can be formally written as
\begin{eqnarray}
&& \frac{d I}{d\tau} + \epsilon' I = \frac{3 \epsilon'}{32 \pi} \int d\Omega' {\mathcal F}_{I}(\Omega, \Omega', \alpha, \beta),
\label{EVS1}\\
&& \frac{d Q}{d\tau} + \epsilon' Q = \frac{3 \epsilon'}{32 \pi} \int d\Omega' {\mathcal F}_{Q}(\Omega,\Omega',\alpha,\beta),
\label{EVS2}\\
&& \frac{d U}{d\tau} + \epsilon' U = \frac{3 \epsilon'}{32 \pi} \int d\Omega' 
{\mathcal F}_{U}(\Omega,\Omega',\alpha,\beta),
\label{EVS3}\\
&& \frac{d V}{d\tau} + \epsilon' V = \frac{3 \epsilon'}{32 \pi} \int d\Omega' 
{\mathcal F}_{V}(\Omega,\Omega',\alpha,\beta),
\label{EVS4}
\end{eqnarray}
where in all the integrands at the right hand side of Eqs. (\ref{EVS1}), (\ref{EVS2}), (\ref{EVS3}) and (\ref{EVS4})
the matrix elements  ${\mathcal T}_{ij}(\Omega, \Omega',\alpha,\beta)$ are functions of, both, the angles of the incident radiation 
$\Omega' = (\vartheta',\varphi')$, the angles of the scattered radiation $\Omega = (\vartheta,\varphi)$ and the 
orientation of the magnetic field defined by the angles $\alpha$ and $\beta$. 
The explicit relations between ${\mathcal T}_{ij}(\Omega,\Omega',\alpha,\beta)$ and the matrix elements $M_{ij}(\Omega,\Omega',\alpha,\beta)$ are reported in the appendix \ref{APPA}. 

\renewcommand{\theequation}{3.\arabic{equation}}
\setcounter{equation}{0}
\section{Brightness perturbations}
\label{sec3}
The brightness perturbations, i.e. the fluctuations of the Stokes parameters in comparison to their equilibrium values
can be decomposed as 
\begin{equation}
 \Delta_{X}(\vec{x},\tau) = \Delta^{(\mathrm{s})}_{X}(\vec{x},\tau) + \Delta^{(\mathrm{v})}_{X}(\vec{x},\tau) + \Delta^{(\mathrm{t})}_{X}(\vec{x},\tau),
\label{BRDEC1}
\end{equation}
where $X=I,\,Q,\,U,\,V$ denotes, generically, one of the four Stokes parameters and where the superscripts refer, respectively, to the scalar, vector and tensor modes of the geometry. The scalar, vector and tensor components of the brightness perturbations are affected, respectively, by the scalar, vector and tensor inhomogeneties of the geometry and of the various sources. Assuming the conformally flat background introduced in 
Eq. (\ref{metric}), the fluctuations of the metric can be written, in general terms, as 
\begin{equation}
\delta g_{\mu\nu}(\vec{x},\tau) = \delta_{\mathrm{s}} g_{\mu\nu}(\vec{x},\tau) 
+ \delta_{\mathrm{v}} g_{\mu\nu}(\vec{x},\tau) + 
 \delta_{\mathrm{t}} g_{\mu\nu}(\vec{x},\tau),
\label{BRDEC1a}
\end{equation}
where $\delta_{\mathrm{s}}$, $\delta_{\mathrm{v}}$ and $\delta_{\mathrm{t}}$ denote 
the inhomogeneity preserving, separately, the scalar, vector and tensor nature of the fluctuations. The scalar modes of the geometry are parametrized in terms of four independent functions $\psi(\vec{x},\tau)$, $\phi(\vec{x},\tau)$, $E(\vec{x},\tau)$ and $F(\vec{x},\tau)$: 
\begin{eqnarray}
&& \delta_{\mathrm{s}} g_{00}(\vec{x},\tau) = 2 a^2(\tau) \phi(\vec{x},\tau), 
\nonumber\\
&& \delta_{\mathrm{s}} g_{0i}(\vec{x},\tau) = - a^2(\tau) \partial_{i} F(\vec{x},\tau),
\nonumber\\
&& \delta_{\mathrm{s}} g_{ij}(\vec{x},\tau) = 2 a^2(\tau) [\psi(\vec{x},\tau) \delta_{ij} - \partial_{i}\partial_{j}E(\vec{x},\tau)].
\label{BRDEC2}
\end{eqnarray}
By setting $E$ and $F$ to zero the gauge freedom is completely fixed and this choice pins down the longitudinal 
(or conformally Newtonian) gauge. 
The vector modes are described by two independent vectors $Q_{i}(\vec{x},\tau)$ and $W_{i}(\vec{x},\tau)$ 
\begin{equation}
 \delta_{\mathrm{v}} g_{0i}(\vec{x},\tau) = - a^2 Q_{i}(\vec{x},\tau),\qquad \delta_{\mathrm{v}} g_{ij}(\vec{x},\tau) = a^2 \biggl[\partial_{i} W_{j}(\vec{x},\tau) + \partial_{j}W_{i}(\vec{x},\tau)\biggr],
\label{BRDEC3}
\end{equation}
subjected to the conditions $\partial_{i} Q^{i} =0$ and $\partial_{i} W^{i} =0$. It will be convenient, for the present purposes, to choose the gauge $Q_{i}=0$. The tensor  modes of the geometry are 
parametrized in terms of a rank-two tensor in three spatial dimensions, i.e.
\begin{equation}
\delta_{t} g_{ij}(\vec{x},\tau) = - a^2 h_{ij}, \qquad \partial_{i} h^{i}_{j}(\vec{x},\tau) = h_{i}^{i}(\vec{x},\tau) = 0,
\label{BRDEC4}
\end{equation}
which is automatically invariant under infinitesimal coordinate transformations.  The following shorthand notation\footnote{The partial derivations with respect to $\tau$ will be denotes by 
$\partial_{\tau}$; the partial derivations with respect to the spatial coordinates 
will be instead denoted by $\partial_{i}$ with $i= 1,\,2,\,3$.} will be adopted
\begin{eqnarray}
&& {\mathcal L}_{I}^{(\mathrm{s})}(\hat{n},\vec{x},\tau) = \partial_{\tau} \Delta^{(\mathrm{s})}_{I}  + \hat{n}^{i} \partial_{i} \Delta^{(\mathrm{s})}_{I} + \epsilon' \Delta^{(\mathrm{s})}_{I} + \frac{1}{q} \biggl(\frac{d q}{d\tau}\biggr)_{\mathrm{s}},
\label{BRDEC6}\\
&& {\mathcal L}_{I}^{(\mathrm{v})}(\hat{n}, \vec{x},\tau) = \partial_{\tau} \Delta^{(\mathrm{v})}_{I} + \hat{n}^{i} \partial_{i} \Delta^{(\mathrm{v})}_{I} + \epsilon' \Delta^{(\mathrm{v})}_{I} + \frac{1}{q}\biggl(\frac{d q}{d\tau}\biggr)_{\mathrm{v}},
\label{BRDEC7}\\
&&  {\mathcal L}_{I}^{(\mathrm{t})}(\hat{n},\vec{x},\tau) = \partial_{\tau} \Delta^{(\mathrm{t})}_{I} + \hat{n}^{i} \partial_{i} \Delta^{(\mathrm{t})}_{I} + \epsilon' \Delta^{(\mathrm{t})}_{I} + \frac{1}{q} \biggl(\frac{d q}{d\tau}\biggr)_{\mathrm{t}},
\label{BRDEC8}
\end{eqnarray}
where $q = \hat{n}_{i} q^{i}$ and where the scalar, vector and tensor contributions to the derivatives of the modulus of the comoving three-momentum are given, respectively, by 
\begin{eqnarray}
&& \biggl(\frac{d q}{d\tau}\biggr)_{\mathrm{s}} = - q \partial_{\tau} \psi + q \hat{n}^{i} \partial_{i} \phi, 
\label{BRDEC9a}\\
&& \biggl(\frac{d q}{d\tau}\biggr)_{\mathrm{v}} = \frac{q}{2} \hat{n}^{i} \hat{n}^{j} (\partial_{i} \partial_{\tau}W_{j} + \partial_{\tau}\partial_{j} W_{i}),
\label{BRDEC9}\\
&& \biggl(\frac{d q}{d\tau}\biggr)_{\mathrm{t}} = - \frac{q}{2} \, \hat{n}^{i}\, \hat{n}^{j} \, \partial_{\tau}h_{ij}.
\label{BRDEC10}
\end{eqnarray}
The identities of Eqs. (\ref{BRDEC9a}), (\ref{BRDEC9}) and (\ref{BRDEC10}) can be derived from the inhomogeneities of Eqs. (\ref{BRDEC2})--(\ref{BRDEC4}) by recalling the definition of comoving three momentum (see discussion after 
Eq. (\ref{metric})) and by using the relations 
\begin{equation}
\frac{d x^{i}}{d\tau} = \frac{P^{i}}{P^{0}} = \frac{q^{i}}{q} = \hat{n}^{i}, 
\label{BRDEC11}
\end{equation}
where $P^{i}$ and $P^{0}$ are the space-like and time-like components of the canonical momentum obeying, for the photons, 
 $g_{\alpha\beta} P^{\alpha} P^{\beta} =0$. 
 The notation introduced in Eqs. (\ref{BRDEC6}), 
 (\ref{BRDEC7}), (\ref{BRDEC8}) for the fluctuations of the intensity can also 
 be generalized to the linear and circular polarizations: 
 \begin{equation}
 {\mathcal L}^{(\mathrm{y})}_{X}(\hat{n},\vec{x},\tau) = \partial_{\tau} \Delta^{(\mathrm{y})}_{X} + \hat{n}^{i} \partial_{i} \Delta^{(\mathrm{y})}_{X}  + \epsilon' \Delta^{(\mathrm{y})}_{X},
\label{BRDEC11a}
\end{equation}
where the subscript can coincide, alternatively, with $Q$, $U$ and $V$ (i.e. $X= Q,\, U,\, V$) and the superscript denotes the 
transformation properties of the given fluctuation (i.e. $\mathrm{y} = \mathrm{s},\,\mathrm{v},\,\mathrm{t}$). The 
fluctuations of the geometry would seem to affect only the brightness perturbation for the intensity but such a conclusion would be incorrect: 
in the presence of a magnetic field the evolution equations of the four brightness 
perturbations are all coupled by the collision term which does not only contain the intensity of the radiation field 
but a weighted sum of the four brightness perturbations integrated  over the directions of the incident radiation. Consequently the polarization of the metric fluctuations 
will also impact on all the four brightness perturbations. 
The conventions on the Fourier transform and polarizations of the scalar, vector and tensor modes will be, in short, 
\begin{eqnarray}
\phi(\vec{x},\tau) &=& \frac{1}{(2\pi)^{3/2}} \int \,d^{3} k \, \phi(\vec{k}, \tau) \,e^{i \vec{k}\cdot \vec{x}},
\label{BRDEC12a}\\
\psi(\vec{x},\tau) &=& \frac{1}{(2\pi)^{3/2}} \int\, d^{3} k \, \psi(\vec{k}, \tau) \,e^{i \vec{k}\cdot \vec{x}},
\label{BRDEC12}\\
W_{i}(\vec{x},\tau) &=&  \frac{1}{(2\pi)^{3/2}} \int \,d^{3} k W_{i}(\vec{k},\tau)\,e^{i \vec{k} \cdot \vec{x}},\qquad \partial_{i} W^{i}(\vec{x},\tau) =0, 
\label{BRDEC13}\\
h_{ij}(\vec{x},\tau) &=& \frac{1}{(2\pi)^{3/2}} \int \,d^{3} k h_{ij}(\vec{k},\tau)\, e^{i \vec{k} \cdot \vec{x}}, \qquad \partial_{i} h^{i}_{j}(\vec{x},\tau) = h_{i}^{i}(\vec{x},\tau) =0.
\label{BRDEC14}
\end{eqnarray}
The vector and the tensor polarizations can be decomposed, respectively, as
\begin{eqnarray}
&& W_{i}(\vec{k},\tau) = \sum_{\lambda} e^{(\lambda)}_{i} W_{(\lambda)}(\vec{k},\tau) = \hat{a}_{i} W_{a}(\vec{k},\tau) + 
\hat{b}_{i} W_{b}(\vec{k},\tau),
\label{BRDEC14a}\\
&& h_{ij}(\vec{k},\tau) =\sum_{\lambda} \epsilon^{(\lambda)}_{ij} h_{(\lambda)}(\vec{k},\tau)  = \epsilon^{\oplus}_{ij} 
h_{\oplus}(\vec{k},\tau) + \epsilon^{\otimes}_{ij} h_{\otimes}(\vec{k},\tau),
\label{BRDEC14b}
\end{eqnarray}
where $\hat{k}$ denotes the direction of propagation and
 the two orthogonal directions $\hat{a}$ and $\hat{b}$ are such that $\hat{a}\times \hat{b} = \hat{k}$. Supposing that the direction of propagation of the relic tensor is 
 oriented along $\hat{k}$,  
the two tensor polarizations are defined in terms of $\hat{a}_{i}$ and $\hat{b}_{i}$ as:
\begin{equation}
\epsilon^{\oplus}_{ij}(\hat{k}) = \hat{a}_{i} \hat{a}_{j} - \hat{b}_{i} \hat{b}_{j}, \qquad 
\epsilon^{\otimes}_{ij}(\hat{k}) = \hat{a}_{i} \hat{b}_{j} + \hat{a}_{j} \hat{b}_{i}.
\label{BRDEC17}
\end{equation}
The projections of the vector and of the tensor polarizations on the direction of photon propagation $\hat{n}$ are:
\begin{eqnarray}
&&\hat{n}^{i} W_{i}(\vec{k},\tau)=  \biggl[ \hat{n}^{i} \hat{a}_{i} W_{a}(\vec{k},\tau) + \hat{n}^{i}\hat{b}_{i} W_{b}(\vec{k},\tau)\biggr],
\label{BRDEC15}\\
&&\hat{n}^{i} \hat{n}^{j} h_{i j}(\vec{k},\tau) =  \biggl\{ [ (\hat{n}\cdot\hat{a})^2 
- (\hat{n}\cdot\hat{b})^2] h_{\oplus}(\vec{k},\tau) + 2 (\hat{n}\cdot\hat{a}) (\hat{n} \cdot \hat{b})
h_{\otimes}(\vec{k},\tau)\biggr\}.
\label{BRDEC16}
\end{eqnarray}
Choosing the direction of propagation of the relic vector and of the relic tensor 
along the $\hat{z}$ axis, the unit vectors 
$\hat{a}$ and $\hat{b}$ will coincide with the remaining two Cartesian directions
and the related Fourier amplitudes will satisfy 
\begin{eqnarray}
&& \hat{n}^{i} W_{i}(\vec{k},\tau) = \sqrt{\frac{2\pi}{3}} \biggl[ W_{L}(\vec{k},\tau) \, Y_{1}^{-1}(\vartheta, \varphi) - W_{R}(\vec{k},\tau) Y_{1}^{1}(\vartheta, \varphi)\biggr] ,
\label{BRDEC18}\\
&& \hat{n}^{i} \hat{n}^{j} h_{i j}(\vec{k},\tau) =  \biggl[ h_{R}(\vec{k},\tau) 
Y_{2}^{2}(\vartheta, \varphi) +  h_{L}(\vec{k},\tau) Y_{2}^{-2}(\vartheta, \varphi)\biggr],
\label{BRDEC19}
\end{eqnarray}
where 
\begin{eqnarray}
&& W_{L}(\vec{k},\tau) = \frac{W_{a}(\vec{k},\tau) + i W_{b}(\vec{k},\tau)}{\sqrt{2}},\qquad W_{R}(\vec{k},\tau) = \frac{W_{a}(\vec{k},\tau) - i W_{b}(\vec{k},\tau)}{\sqrt{2}},
\nonumber\\
&& h_{L}(\vec{k},\tau) = \frac{h_{\oplus}(\vec{k},\tau) + i h_{\otimes}(\vec{k},\tau)}{\sqrt{2}},\qquad h_{R}(\vec{k},\tau) = \frac{h_{\oplus}(\vec{k},\tau) - i h_{\otimes}(\vec{k},\tau)}{\sqrt{2}};
\label{BRDEC20}
\end{eqnarray}
the spherical harmonics appearing in Eqs. (\ref{BRDEC18}) and (\ref{BRDEC19}) 
are, respectively, 
\begin{equation}
Y_{1}^{\pm 1} (\vartheta,\varphi) = \mp \sqrt{\frac{3}{8\pi}} \, \sin{\vartheta} \,\,e^{\pm i \varphi},\qquad Y_{2}^{\pm 2} (\vartheta,\varphi) = \sqrt{\frac{15}{32 \pi}} \sin^2{\vartheta} \,\,e^{ \pm 2 i \varphi},
\label{BRDEC21}
\end{equation}
showing, as well known in the context of the total angular momentum method \cite{total}, that the vector and tensor modes 
excite, respectively, the two harmonics given in Eq. (\ref{BRDEC21}). While the total angular momentum method can be generalized to the case of an arbitrarily oriented magnetic field, we prefer to work, in the present context, with the formalism which is more directly applicable to numerical codes and to standard analytic estimates. 
In Fourier space  Eqs. (\ref{BRDEC6}), (\ref{BRDEC7}) and (\ref{BRDEC8}) become
\begin{eqnarray}
 {\mathcal L}_{I}^{(\mathrm{s})}(\mu, \varphi,\vec{k},\tau) &=& \partial_{\tau}\Delta^{(\mathrm{s})}_{I} + (i k \mu + \epsilon') \Delta^{(\mathrm{s})}_{I}  + i k\mu \phi  - \partial_{\tau} \psi,
\label{BRDEC6a}\\
 {\mathcal L}_{I}^{(\mathrm{v})}(\mu,\varphi,\vec{k},\tau) &=& \partial_{\tau} \Delta^{(\mathrm{v})}_{I} +  (i k \mu + \epsilon')\Delta^{(\mathrm{v})}_{I}
 \nonumber\\
 &+& \sqrt{\frac{2\pi}{3}} i\, \mu\,\biggl[ \partial_{\tau} W_{L}(\vec{k},\tau) \, Y_{1}^{-1}(\vartheta, \varphi) - \partial_{\tau} W_{R}(\vec{k},\tau) Y_{1}^{1}(\vartheta, \varphi)\biggr],
\label{BRDEC7a}\\
 {\mathcal L}_{I}^{(\mathrm{t})}(\mu,\varphi,\vec{k},\tau) &=& \partial_{\tau}\Delta^{(\mathrm{t})}_{I} + (i k \mu + \epsilon') \Delta^{(\mathrm{t})}_{I} 
 \nonumber\\
&-& \sqrt{\frac{2\pi}{15}} \biggl[ \partial_{\tau}h_{R}(\vec{k},\tau) 
Y_{2}^{2}(\vartheta, \varphi) +  \partial_{\tau} h_{L}(\vec{k},\tau) Y_{2}^{-2}(\vartheta, \varphi)\biggr].
\label{BRDEC8a}
\end{eqnarray}
Similarly Eq. (\ref{BRDEC11a}) becomes, in Fourier space,   
\begin{equation}
{\mathcal L}^{(\mathrm{y})}_{X}(\mu,\varphi,\vec{k},\tau) = \partial_{\tau} \Delta^{(\mathrm{y})}_{X}+ ( i k \mu +\epsilon') \Delta^{(\mathrm{y})}_{X}. 
\label{BRDEC11b}
\end{equation}
The explicit form of the transport equations for the scalar, vector and tensor modes of the  geometry will be scrutinized in the three forthcoming sections. 
\renewcommand{\theequation}{4.\arabic{equation}}
\setcounter{equation}{0}
\section{Scalar modes}
\label{sec4}
Following the notation of Eqs. (\ref{BRDEC6a}) and (\ref{BRDEC11b}) the scalar transport equations can be formally expressed as 
\begin{eqnarray}
&& {\mathcal L}^{(\mathrm{s})}_{I}(\mu,\varphi,\vec{k},\tau) =  \epsilon' \hat{n}^{i} v^{(\mathrm{s})}_{i} + \frac{3 \epsilon'}{32\pi} \int_{-1}^{1} \,d\nu \int_{0}^{2\pi}d\varphi'{\mathcal F}^{(\mathrm{s})}_{I}(\mu,\nu, \varphi,\varphi', \alpha,\beta),
\label{S1}\\
&&  {\mathcal L}^{(\mathrm{s})}_{Q}(\mu,\varphi,\vec{k},\tau) = \frac{3 \epsilon'}{32 \pi}
\int_{-1}^{1}d\nu \int_{0}^{2 \pi}\, d\varphi'{\mathcal F}^{(\mathrm{s})}_{\mathrm{Q}}(\mu, \nu,\varphi, \varphi',\alpha,\beta),
\label{S2}\\
&& {\mathcal L}^{(\mathrm{s})}_{U}(\mu,\varphi,\vec{k},\tau) =   \frac{3 \epsilon'}{32 \pi}
\int_{-1}^{1} \,d\nu \int_{0}^{2\pi}d\varphi'\,\, {\mathcal F}^{(\mathrm{s})}_{\mathrm{U}}(\mu,\nu,\varphi,\varphi',\alpha,\beta),
\label{S3}\\
&& {\mathcal L}^{(\mathrm{s})}_{V}(\mu,\varphi,\vec{k},\tau) =  \frac{3 \epsilon'}{32 \pi}
\int_{-1}^{1}d\nu \int_{0}^{2\pi}d\varphi'\,\,{\mathcal F}^{(\mathrm{s})}_{\mathrm{V}}(\mu,\nu,\varphi,\varphi',\alpha,\beta),
\label{S4}
\end{eqnarray}
where $v^{(\mathrm{s})}_{i}$ denotes the scalar component of the baryon velocity field. 
In Eqs. (\ref{S1})--(\ref{S4}) the source terms involve the integration over the incoming photon directions. Both the integration over $\varphi'$ and $\nu$  
can be performed explicitly and the final expressions are rather lengthy, as easily 
imaginable. To make the explicit equations more manageable without loosing 
any relevant information it is useful, in the following part of the present section, to write the results already in the physical limit, i.e. 
 owing to the numerical values of the plasma and Larmor frequencies and recalling 
 Eq. (\ref{LAM3})
\begin{eqnarray}
&& \Lambda_{1}(\omega) = \Lambda_{2}(\omega) = \Lambda_{3}(\omega)  = 1 + {\mathcal O}(m_{\mathrm{e}}/m_{\mathrm{p}}),
\nonumber\\ 
&& \zeta(\omega) \simeq -1 + f_{\mathrm{e}}^2(\omega) + {\mathcal O}(f_{\mathrm{e}}^4).
\label{SC0}
\end{eqnarray}
The scalar source terms depend upon the explicit form of the matrix elements appearing in Eqs. (\ref{FI}), (\ref{FQ}), (\ref{FU}) and (\ref{FV}). The integration over $\varphi'$ can be performed explicitly. Using the notation 
\begin{equation}
\overline{{\mathcal T}}_{ab}(\mu,\nu, \varphi, \alpha,\beta) = 
\int_{0}^{2\pi} d\varphi' \,{\mathcal T}_{a b}(\mu,\nu, \varphi, \varphi',\alpha,\beta),
\label{SC1}
\end{equation}
the final results are reported, for completeness and future peruse, in appendix \ref{APPB}.  According to Eqs. (\ref{S1})--(\ref{S4}), 
the expressions reported in Eqs. (\ref{TbarII})-(\ref{TbarVV}) must be integrated over $\nu$. For the $\nu$ integration 
it is useful to expand the various brightness perturbations in a series of Legendre polynomials $P_{\ell}(\nu)$
\begin{equation}
\Delta_{X}(\nu,k,\tau) = \sum_{\ell} (-i)^{\ell} (2\ell + 1) \, P_{\ell}(\nu)\, \Delta_{X\,\ell}(k,\tau).
\label{r1}
\end{equation}
The integration over $\nu$ will then have the net result of expressing the source terms in terms of a limited 
number of multipoles of the intensity and of the polarization. In explicit terms the source terms can be expressed, for each 
brightness perturbation, as an expansion in $f_{\mathrm{e}}(\omega)$:
\begin{eqnarray}
&& \partial_{\tau} \Delta^{(\mathrm{s})}_{I}
+ ( i k\mu + \epsilon') \Delta^{(\mathrm{s})}_{I} = \partial_{\tau} \psi - i k \mu \phi + \epsilon' {\mathcal A}_{I} + \epsilon' \, f_{\mathrm{e}}(\omega) \, 
{\mathcal B}_{I} + \epsilon' \, f_{\mathrm{e}}^2(\omega) \, {\mathcal C}_{I}, 
\label{BRI}\\
&&\partial_{\tau} \Delta^{(\mathrm{s})}_{Q} + ( i k\mu + \epsilon') \Delta^{(\mathrm{s})}_{Q} =  \epsilon' {\mathcal A}_{Q} + \epsilon' \, f_{\mathrm{e}}(\omega) \, 
{\mathcal B}_{Q} + \epsilon' \, f_{\mathrm{e}}^2(\omega) \, {\mathcal C}_{Q},
\label{BRQ}\\
&&   \partial_{\tau} \Delta^{(\mathrm{s})}_{U} + ( i k\mu + \epsilon') \Delta^{(\mathrm{s})}_{U} =  \epsilon' {\mathcal A}_{U} + \epsilon' \, f_{\mathrm{e}}(\omega) \, 
{\mathcal B}_{U} + \epsilon' \, f_{\mathrm{e}}^2(\omega) \, {\mathcal C}_{U},
\label{BRU}\\
&&  \partial_{\tau} \Delta^{(\mathrm{s})}_{V}+ ( i k\mu + \epsilon') \Delta^{(\mathrm{s})}_{V} =  \epsilon' {\mathcal A}_{V} + \epsilon' \, f_{\mathrm{e}}(\omega) \, 
{\mathcal B}_{V} + \epsilon' \, f_{\mathrm{e}}^2(\omega) \, {\mathcal C}_{V},
\label{BRV}
\end{eqnarray}
where, for $X= I,\,Q,\, U,\, V$, 
${\mathcal A}_{X}$ denotes the leading order result, ${\mathcal B}_{X}$ denotes the next-to-leading order (NLO) correction while 
${\mathcal C}_{X}$ denotes the next-to-next-to-leading (NNLO) term.
Defining with $S_{P}$ the usual combination of the quadrupole of the intensity 
and of the monopole and quadrupole of the linear polarization 
(see, e.g. \cite{mab,sel1,sel2})
\begin{equation}
S_{P} = \Delta_{I 2} + \Delta_{Q 0} + \Delta_{Q 2}, 
\label{SP}
\end{equation}
the  leading order contribution for the for brightness perturbations is then given by:
\begin{eqnarray}
{\mathcal A}_{I} &=& \Delta_{I 0} + \mu v_{\mathrm{b}} - \frac{P_{2}(\mu)}{2} S_{P},
\label{AI}\\
{\mathcal A}_{Q} &=& \frac{3}{4} (1 - \mu^2) S_{P},
\label{AQ}\\
{\mathcal A}_{U} &=& 0,\qquad {\mathcal A}_{V} = - \frac{3}{2} \, i \,\mu\, \Delta_{V 1},
\label{AVAU}
\end{eqnarray}
where the notation $\vec{v}^{(\mathrm{s})}= \vec{k} v_{\mathrm{b}}$ has been 
employed for the scalar component of the Doppler term.
The NLO contribution to the right hand side of Eqs. (\ref{BRI}), 
(\ref{BRQ}), (\ref{BRU}) and (\ref{BRV}) is 
\begin{eqnarray}
{\mathcal B}_{I} &=& - \frac{3}{2} \,i \,\biggl[ (1 + \mu^2) \cos{\beta} + \mu \sqrt{1 - \mu^2} \cos{(\varphi - \alpha)} \sin{\beta} \biggr] \Delta_{V 1}
\label{BI}\\
{\mathcal B}_{Q} &=& - \frac{3}{2} \, i \, \Delta_{V 1} \biggl[ (\mu^2 -1) \cos{\beta} + \mu \sqrt{1 - \mu^2} \cos{(\varphi - \alpha)} \sin{\beta} \biggr],
\label{BQ}\\
{\mathcal B}_{U} &=& \frac{3}{2}\, i\, \Delta_{V 1} \sqrt{ 1 -\mu^2} \sin{\beta} \sin{(\varphi - \alpha)},
\label{BU}\\
{\mathcal B}_{V} &=& \biggl[ \mu \cos{\beta} - \frac{\sqrt{1 - \mu^2}}{2} \cos{(\varphi - \alpha)} \sin{\beta} \biggr] \Delta_{I 0} 
\nonumber\\
&-& \biggl[ 2 \mu \cos{(\varphi - \alpha)} \sin^2{\beta} + \frac{\sqrt{ 1 - \mu^2}}{2} \sin{\beta} \biggl] \Delta_{I 2} 
\nonumber\\
&-& \biggl[ \frac{\mu}{2} \cos{\beta}  - \frac{\sqrt{ 1 - \mu^2}}{4} \cos{(\varphi - \alpha)} \sin{\beta} \biggr] (\Delta_{Q 2} + 
\Delta_{Q 0}).
\label{BV}
\end{eqnarray}
Finally, the NNLO contribution to the right hand side of Eqs. (\ref{BRI}), (\ref{BRQ}), (\ref{BRU}) and (\ref{BRV}) is
\begin{eqnarray}
{\mathcal C}_{I} &=& \biggl[ (\mu^2 + 1) + \mu \sqrt{1 - \mu^2} \cos{(\varphi - \alpha)} \sin{2 \beta}\biggr]\biggl( \frac{\Delta_{I 0}}{2}
 - \Delta_{I 2}\biggr) 
 \nonumber\\
&+& \frac{1}{2} \biggl[ \mu^2 ( \Delta_{I 0} + \Delta_{I 2}) + (\Delta_{I 2} - 2 \Delta_{I 0}) \biggr] \sin^2{\beta} 
 \cos^2{(\varphi - \alpha)}  -\frac{\cos{2\beta}}{2} ( 1 + 3 \mu^2 )  
 \nonumber\\
 &-& \biggl\{ (1 - \mu^2) \sin^2{(\varphi - \alpha)}  + \frac{1 + \mu^2}{4} \biggl[ \cos{2 (\varphi - \alpha - \beta)} + 
  \cos{2 (\varphi - \alpha + \beta)} \biggr] 
 \nonumber\\ 
&-& \frac{\mu \sqrt{1 - \mu^2}}{2} \biggl[ \sin{(\varphi - \alpha - 2\beta)} - \sin{(\varphi - \alpha + 2 \beta)}\biggr] \biggr\} (\Delta_{Q 0} + \Delta_{Q 2}),
\label{CI}\\
{\mathcal C}_{Q} &=& \biggl\{ \frac{\mu^2 -1}{2} + \frac{\sin^2{\beta}}{8} \biggl[ 4 - 2 ( 2 \mu^2 + 1) \cos^2{(\varphi - \alpha)}\biggr] 
+ \frac{ \mu \sqrt{ 1 - \mu^2 }}{2} \cos{(\varphi - \alpha)} \sin{2 \beta}\biggr\} \Delta_{I 0}
\nonumber\\
&-& \biggl\{ \frac{\mu^2 -1}{4} + \frac{\sin^2{\beta}}{2}\biggl[ 1 + (1 - \mu^2) \cos^2{(\varphi - \alpha)}\biggr] 
+ \frac{\mu \sqrt{ 1 - \mu^2}}{4} \cos{(\varphi - \alpha)} \sin{2 \beta} \biggr\} \Delta_{I 2} 
\nonumber\\
&+& \biggl\{ \frac{\mu^2 -1}{8} - \frac{\mu^2 + 1}{8} \cos{ 2(\varphi - \alpha)} + \frac{3}{8} ( 1 - \mu^2) \cos{2\beta}
\nonumber\\
&+& \frac{\mu^2 + 1}{16} \biggl[ \cos{2 (\varphi - \alpha - \beta)} + \cos{ 2 (\varphi - \alpha + \beta)}\biggr]
\nonumber\\
&+& \frac{ \mu \sqrt{1 - \mu^2}}{8} \biggl[ \sin{(\varphi -\alpha - 2 \beta)} - \sin{(\varphi - \alpha + 2 \beta)}\biggr]\biggr\} (\Delta_{Q 0} 
+ \Delta_{Q 2}),
\label{CQ}\\
{\mathcal C}_{U} &=& 
\sin{(\varphi - \alpha)} \biggl[ \frac{\sqrt{ 1 - \mu^2}}{2} \sin{2 \beta} - \mu \cos{(\varphi - \alpha)} \sin^2{\beta}\biggr]
 \Delta_{I 0} 
 \nonumber\\
 &+& \sin{(\varphi - \alpha)} \biggl[ \mu \cos{(\varphi - \alpha)} \sin^2{\beta} - \frac{\sqrt{1 - \mu^2}}{4} \sin{2 \beta}\biggr] 
 \Delta_{I 2} 
 \nonumber\\
 &+& \frac{\sin{\beta} \sin{(\varphi - \alpha)}}{2} \biggl[ \sqrt{ 1 - \mu^2 } \cos{\beta} + 2 \mu \cos{(\varphi -\alpha)} \sin{\beta}\biggr]
 (\Delta_{Q 2} + \Delta_{Q 0}),
 \label{CU}\\
{\mathcal C}_{V} &=& - \frac{3}{2} \, i\, \biggl[ \mu \cos^2{\beta} + \sqrt{ 1 - \mu^2 } \cos{(\varphi -\alpha)} \sin{\beta} \cos{\beta}\biggr]
\Delta_{V 1}.
\label{CV}
\end{eqnarray}
Several cross-checks on the obtained results have been made; they will be swiftly 
mentioned and can be directly reproduced by using the results 
reported in the appendices \ref{APPA} and \ref{APPB}:
\begin{itemize}
\item{} it has been verified explicitly at the level of the exact expressions (i.e. without implementing 
the limit of Eq. (\ref{SC0})) the equations must be independent upon the 
$\alpha$ and $\beta$ once $(\hat{e}_{3}\cdot\vec{B})\to 0$: this is exactly 
what happens;
\item{} it has been verified that in the limit $\alpha = \beta =0$ the exact expressions must reproduce the partial results already obtained in \cite{mgpol2,mgpol4,mgpol5}; with Eqs. (\ref{BRI})--(\ref{BRV}) few typos present in the published version of \cite{mgpol2} are corrected;
\item{} by averaging of the source terms over $\alpha$ and $\beta$ 
terms proportional to $f_{\mathrm{e}}(\omega)$ should automatically 
disappear without performing any specific limit: this is what will be 
explicitly shown in the remaining part of this section.
\end{itemize}
The remaining part of the section is devoted to the averaging of the source 
terms over the magnetic field directions as suggested in the last point of the above list of items. By integrating over $\alpha$ and $\beta$ the 
source functions appearing at the right hand side of 
Eqs. (\ref{S1}), (\ref{S2}), (\ref{S3}) and (\ref{S4}), the evolution equations for the brightness perturbations read
\begin{eqnarray}
&& {\mathcal L}^{(\mathrm{s})}_{I}(\mu,\varphi,\vec{k},\tau) =  \epsilon' \hat{n}^{i} v^{(\mathrm{s})}_{i} 
\nonumber\\
&&+ \frac{3 \epsilon'}{128\pi^2} \int_{-1}^{1}d\nu \int_{0}^{2\pi}d\varphi' \int_{0}^{\pi} \sin{\beta}d\beta \int_{0}^{2\pi}d\alpha {\mathcal F}^{(\mathrm{s})}_{I}(\mu,\nu, \varphi,\varphi', \alpha,\beta),
\label{S1a}\\
&&  {\mathcal L}^{(\mathrm{s})}_{Q}(\mu,\varphi,\vec{k},\tau) = \frac{3 \epsilon'}{128 \pi^2}
\int_{-1}^{1}\,d\nu \int_{0}^{2 \pi} d\varphi'\int_{0}^{\pi} \sin{\beta}d \beta \int_{0}^{2\pi}d\alpha{\mathcal F}^{(\mathrm{s})}_{\mathrm{Q}}(\mu, \nu,\varphi, \varphi',\alpha,\beta),
\label{S2a}\\
&& {\mathcal L}^{(\mathrm{s})}_{U}(\mu,\varphi,\vec{k},\tau) =   \frac{3 \epsilon'}{128 \pi^2}
\int_{-1}^{1} d\nu \int_{0}^{2\pi}d\varphi'\, \int_{0}^{\pi} \sin{\beta}d \beta \int_{0}^{2\pi} d\alpha {\mathcal F}^{(\mathrm{s})}_{\mathrm{U}}(\mu,\nu,\varphi,\varphi',\alpha,\beta),
\label{S3a}\\
&& {\mathcal L}^{(\mathrm{s})}_{V}(\mu,\varphi,\vec{k},\tau) =  \frac{3 \epsilon'}{128 \pi^2}\int_{-1}^{1} d\nu \int_{0}^{2\pi} d\varphi'\int_{0}^{\pi} \sin{\beta}d \beta 
\int_{0}^{2\pi}d\alpha {\mathcal F}^{(\mathrm{s})}_{\mathrm{V}}(\mu,\nu,\varphi,\varphi',\alpha,\beta),
\label{S4a}
\end{eqnarray}
where the factor $128 \pi^2$ accounts for the $4\pi$ factor arising from the 
average over the solid angle spanned by $\alpha$ and $\beta$.
By performing the averages explicitly, the evolution equations 
of the four brightness perturbations read:
\begin{eqnarray}
 \partial_{\tau} \Delta^{(\mathrm{s})}_{I}
+ ( i k\mu + \epsilon') \Delta^{(\mathrm{s})}_{I} &=& \partial_{\tau} \psi - i k \mu \phi + \epsilon' \biggl[ \Delta_{I 0} + \mu v_{\mathrm{b}} - \frac{P_{2}(\mu)}{2} S_{\mathrm{P}} 
\nonumber\\
&+& f_{\mathrm{e}}^2 \biggl(\frac{2}{3} \Delta_{I 0} + \frac{P_{2}(\mu)}{6} S_{\mathrm{P}}\biggr)\biggr]
\label{AVI}\\
\partial_{\tau} \Delta^{(\mathrm{s})}_{Q} + ( i k\mu + \epsilon') \Delta^{(\mathrm{s})}_{Q} &=&  \epsilon'  \frac{(f_{\mathrm{e}}^2 - 3) (\mu^2 -1)}{4} S_{\mathrm{P}},
\label{AVQ}\\
\partial_{\tau} \Delta^{(\mathrm{s})}_{U} + ( i k\mu + \epsilon') \Delta^{(\mathrm{s})}_{U} &=&  0,
\label{AVU}\\
\partial_{\tau} \Delta^{(\mathrm{s})}_{V}+ ( i k\mu + \epsilon') \Delta^{(\mathrm{s})}_{V} &=& - \frac{i \, \epsilon'}{2}( 3 + f_{\mathrm{e}}^2) \Delta_{V 1}.
\label{AVV}
\end{eqnarray}
The results of the present section make quantitatively clear that 
if the magnetic field has a predominant direction over typical scales 
comparable with the wavelengths of the scattered photons, then 
the circular polarization is larger than in the case where, over the same 
physical scales the magnetic field is randomly oriented.  It has been argued 
in \cite{mgpol1,mgpol2} (see also \cite{mgpol5}) that over small angular scales 
the maximal amount of circular polarization arises when there is a strong 
alignment of the magnetic field along the direction of propagation of the photon 
which coincides, for large multipoles, with the third Cartesian direction. 

The present results improve and confirm, at once, the assumptions 
made in the analytic and numerical estimates of magnetized CMB anisotropies of 
Refs. \cite{mg1,mg2,mg3}.  Indeed, the scalar fluctuations 
of the geometry obey a set of evolution equations where 
large-scale magnetic fields contribute in many respects. 
These equations will not be repeated here and can be 
found in \cite{mg1,mg2,mg3}. The present results 
improve on the transport equations used there and pave 
the way for a more consistent account the effects of 
pre-decoupling magnetic fields both at large as well as at small
angular scales. 
\renewcommand{\theequation}{5.\arabic{equation}}
\setcounter{equation}{0}
\section{Vector modes}
\label{sec5}
In the case of the vector modes of the geometry 
the integration over $\varphi'$ of the source functions cannot be easily performed as in the case of the scalar modes of the geometry
(see section \ref{sec4}). Each of the two vector polarizations induce a different angular dependence in the corresponding brightness perturbations. In this paper three categories of circular and linear polarizations can be defined:
\begin{itemize}
\item{} the linear and circular polarizations of the scattered (and incident) photons (already described in sections \ref{sec2} and \ref{sec3}) which are described 
by the four Stokes parameters or by the appropriate Stokes matrix;
\item{} the linear and circular polarizations of the relic vector waves 
which we are going to discuss in the present section and which have been 
already introduced, respectively, in Eqs. (\ref{BRDEC15}) and (\ref{BRDEC18});
\item{} the linear and circular polarizations of the relic tensor  
waves introduced, respectively, in Eqs. (\ref{BRDEC16}) and (\ref{BRDEC19}) and 
discussed in the following section \ref{sec4}.
\end{itemize}
The linear and circular polarizations of the relic tensor and vector waves 
are just equivalent basis for the description of the tensor and vector modes 
of the geometry. To avoid potential confusions 
the vector and the tensor waves will always be treated in the basis of the linear polarizations. In full analogy with the treatment of section \ref{sec4}  the evolution equations for the vector components of the brightness perturbations can be formally written as 
\begin{eqnarray}
&& {\mathcal L}^{(\mathrm{v})}_{I}(\mu,\varphi,\vec{k},\tau) =  \epsilon' n^{i} v^{(\mathrm{v})}_{i} + \frac{3 \epsilon'}{32\pi} \int_{-1}^{1} \,d\nu \int_{0}^{2\pi}\, d\varphi'\, {\mathcal F}^{(\mathrm{v})}_{I}(\mu,\nu, \varphi,\varphi', \alpha,\beta),
\label{V1}\\
&&  {\mathcal L}^{(\mathrm{v})}_{Q}(\mu,\varphi,\vec{k},\tau) = \frac{3 \epsilon'}{32 \pi}
\int_{-1}^{1}\,d\nu \int_{0}^{2 \pi}\, d\varphi'\,\,{\mathcal F}^{(\mathrm{v})}_{\mathrm{Q}}(\mu, \nu,\varphi, \varphi',\alpha,\beta),
\label{V2}\\
&& {\mathcal L}^{(\mathrm{v})}_{U}(\mu,\varphi,\vec{k},\tau) =   \frac{3 \epsilon'}{32 \pi}
\int_{-1}^{1} \,d\nu \int_{0}^{2\pi}\, d\varphi'\,\, {\mathcal F}^{(\mathrm{v})}_{\mathrm{U}}(\mu,\nu,\varphi,\varphi',\alpha,\beta),
\label{V3}\\
&& {\mathcal L}^{(\mathrm{v})}_{V}(\mu,\varphi,\vec{k},\tau) =  \frac{3 \epsilon'}{32 \pi}
\int_{-1}^{1} \,d\nu \int_{0}^{2\pi}\, d\varphi'\,\,{\mathcal F}^{(\mathrm{v})}_{\mathrm{V}}(\mu,\nu,\varphi,\varphi',\alpha,\beta),
\label{V4}
\end{eqnarray}
where $ v^{(\mathrm{v})}_{i}$ denotes the vector component of the baryon velocity field. The polarizations of the baryon velocity will follow the same kind of decomposition illustrated for the vector of the geometry in Eq. (\ref{BRDEC18}).
The relative directions of the magnetic field intensity and of the photon propagation determine the polarization of the outgoing radiation. Following the strategy described in section \ref{sec2} the direction 
of propagation of the relic vector wave can be fixed and  the direction of the magnetic field varied at wish.

Consider first the case where the magnetic field is oriented along the same direction of the vector wave and suppose, without loss of generality, that 
the vector propagates along $\hat{k} = \hat{z}$. Since, in this case, $\alpha=\beta=0$  and Eqs. (\ref{DIP4}), (\ref{DIP5}), (\ref{DIP6}) and (\ref{DIP7}) will lead, respectively, to the following matrix elements: 
\begin{eqnarray}
M_{11}(\mu,\varphi,\nu,\varphi') &=& \zeta \mu \nu \Lambda_{1}  \cos{(\varphi' - \varphi)} 
- \sqrt{1 - \mu^2} \sqrt{1 - {\nu}^2} \Lambda_{3}
\nonumber\\
&-& i \Lambda_{2} f_{\mathrm{e}} \zeta \mu\nu \sin{(\varphi' - \varphi)},
\nonumber\\
M_{12}(\mu,\varphi,\nu,\varphi') &=&  - \zeta \mu \Lambda_{1} \sin{\Delta\varphi} - i \Lambda_{2} f_{\mathrm{e}}
 \zeta \mu \cos{(\varphi' - \varphi)},
\nonumber\\
M_{21}(\mu,\varphi,\nu,\varphi') &=& \zeta \nu \Lambda_{1}
 \sin{\Delta\varphi} + i f_{\mathrm{e}} \Lambda_{2} \zeta \nu \cos{(\varphi' - \varphi)},
\nonumber\\
M_{22}(\mu,\varphi,\nu,\varphi') &=& \zeta \Lambda_{1} \cos{(\varphi' - \varphi)} 
- i f_{\mathrm{e}} \Lambda_{2} \zeta \sin{(\varphi' - \varphi)}.
\label{V5}
\end{eqnarray}
Using Eq. (\ref{V5}) the source terms of Eqs. (\ref{V1}), (\ref{V2}), (\ref{V3}) and (\ref{V4}) can be computed. The explicit form of Eqs. (\ref{V1})--(\ref{V4}) is rather lengthy: instead of writing {\em all} the equations,  Eq. (\ref{V1}) will just be written for illustration with the purpose of demonstrating how the different vector polarizations induce a specific azimuthal dependence 
in the vector brightness perturbations. Equation (\ref{V1}) written in the basis of the linear polarizations reads
\begin{eqnarray}
&& \partial_{\tau} \Delta_{I}^{(\mathrm{v})} + ( i k \mu + \epsilon') \Delta_{I}^{(\mathrm{v})} + i\,\mu\,\sqrt{1 - \mu^2} \biggl[ 
\cos{\varphi} \partial_{\tau} W_{a} + \sin{\varphi} \partial_{\tau} W_{b} \biggr]
\nonumber\\
&& = \epsilon' \sqrt{1 - \mu^2} [ \cos{\varphi} v_{a} + \sin{\varphi} v_{b}] +  \frac{3 \epsilon'}{32\pi} \int_{-1}^{1} \,d\nu \int_{0}^{2\pi}\, d\varphi'\, {\mathcal F}^{(\mathrm{v})}_{I}(\mu,\nu, \varphi,\varphi', \alpha,\beta),
\label{V6}
\end{eqnarray}
where, according to Eq. (\ref{FI})  the integrand of the source term acts on the vector components of the various 
brightness perturbations and it is given by 
\begin{eqnarray}
&& {\mathcal F}^{(\mathrm{v})}_{I}(\mu,\nu, \varphi,\varphi', \alpha,\beta) = {\mathcal T}_{II}(\mu,\nu, \varphi,\varphi', \alpha,\beta) 
\Delta^{(\mathrm{v})}_{I}(\nu,\varphi') + 
 {\mathcal T}_{IQ}(\mu,\nu, \varphi,\varphi', \alpha,\beta)\Delta^{(\mathrm{v})}_{Q}(\nu,\varphi') 
\nonumber\\
&& +{\mathcal T}_{IU}(\mu,\nu, \varphi,\varphi', \alpha,\beta) \Delta^{(\mathrm{v})}_{U}(\nu,\varphi') 
+ {\mathcal T}_{IV}(\mu,\nu, \varphi,\varphi', \alpha,\beta) \Delta^{(\mathrm{v})}_{V}(\nu,\varphi').
\label{V7}
\end{eqnarray}
After inspection of all the four expressions appearing in Eqs. (\ref{V1}), (\ref{V2}), (\ref{V3}) and (\ref{V4}) it can be checked that the consistent ansatz for the four brightness perturbations is given by
\begin{eqnarray}
&& \Delta^{(\mathrm{v})}_{I}(\varphi,\mu,k,\tau) = \sqrt{1 - \mu^2} \biggl[ \cos{\varphi} {\mathcal M}_{a}(k,\tau) + \sin{\varphi} {\mathcal M}_{b}(k,\tau)\biggr],
\label{V8}\\
&& \Delta^{(\mathrm{v})}_{Q}(\varphi,\mu,k,\tau) = \mu\sqrt{1 - \mu^2} \biggl[ \cos{\varphi} {\mathcal N}_{a}(k,\tau) + \sin{\varphi} {\mathcal N}_{b}(k,\tau)\biggr],
\label{V9}\\
&& \Delta^{(\mathrm{v})}_{U}(\varphi,\mu,k,\tau) = \sqrt{1 - \mu^2} \biggl[ -\sin{\varphi} {\mathcal N}_{a}(k,\tau) + \cos{\varphi} {\mathcal N}_{b}(k,\tau)\biggr],
\label{V10}\\
&& \Delta^{(\mathrm{v})}_{V}(\varphi,\mu,\tau) = \sqrt{1 - \mu^2} \biggl[ \cos{\varphi} {\mathcal V}_{a}(k,\tau) + \sin{\varphi} {\mathcal V}_{b}(k,\tau)\biggr].
\label{V11}
\end{eqnarray}
The equations obeyed by ${\mathcal M}_{a}$, ${\mathcal N}_{a}$ and ${\mathcal V}_{a}$ are the same 
as the ones obeyed by ${\mathcal M}_{b}$, ${\mathcal N}_{b}$ and ${\mathcal V}_{b}$ and they can be written, for a 
generic linear polarization, as
\begin{eqnarray}
&& \partial_{\tau} {\mathcal M} + ( i k \mu + \epsilon') {\mathcal M} + i \mu \partial_{\tau} W = \epsilon' v +
\epsilon' \mu \zeta \Lambda_{1} \Lambda_{3}  \Sigma_{1}^{(\mathrm{v})} + \epsilon' \mu f_{\mathrm{e}} \zeta \Lambda_{2} \Lambda_{3} \Sigma_{2}^{(\mathrm{v})},
\label{V12}\\
&& \partial_{\tau} {\mathcal N} + ( i k \mu + \epsilon') {\mathcal N}=  \epsilon' \zeta \Lambda_{1} \Lambda_{3}   
\Sigma_{1}^{(\mathrm{v})}  +\epsilon' f_{\mathrm{e}} \Lambda_{2} \Lambda_{3} \Sigma_{2}^{(\mathrm{v})},
\label{V13}\\
&& \partial_{\tau} {\mathcal V} + ( i k \mu + \epsilon') {\mathcal V} = \epsilon' f_{\mathrm{e}} \zeta \Lambda_{2} \Lambda_{3}  \Sigma^{(\mathrm{v})}
+ \epsilon' \zeta \Lambda_{1} \Lambda_{3} \Sigma_{2}^{(\mathrm{v})},
\end{eqnarray}
where the two newly defined source functions $\Sigma_{1}^{(\mathrm{v})}$ and $\Sigma_{2}^{(\mathrm{v})}$ 
are given by:
\begin{eqnarray}
\Sigma_{1}^{(\mathrm{v})}(k,\tau) &=&
 \frac{3}{8} \int_{-1}^{1} \biggl[ \nu (\nu^2 -1) {\mathcal M}(k,\nu,\tau) + (\nu^4 -1) {\mathcal N}(k,\nu,\tau)\biggr]\, d\nu,
\label{V14}\\
\Sigma_{2}^{(\mathrm{v})}(k,\tau) &=&  \frac{3}{8} \int_{-1}^{1} (\nu^2 -1)\, {\mathcal V}(k,\nu,\tau) \, d\nu.
\label{V15}
\end{eqnarray}
The source functions appearing in Eqs. (\ref{V14}) and (\ref{V15}) can be made more explicit by expanding 
${\mathcal M}(\nu,k, \tau)$, ${\mathcal N}(\nu,k,\tau)$ and ${\mathcal V}(\nu,k,\tau)$ with the same conventions employed in 
Eq. (\ref{r1}):
\begin{eqnarray}
{\mathcal M}(\nu,k,\tau)  &=& \sum_{\ell} (-i)^{\ell} (2\ell + 1) \, P_{\ell}(\nu)\, {\mathcal M}_{\ell}(k,\tau).
\label{MM1}\\
{\mathcal N}(\nu,k,\tau)  &=& \sum_{\ell} (-i)^{\ell} (2\ell + 1) \, P_{\ell}(\nu)\, {\mathcal N}_{\ell}(k,\tau).
\label{NN1}\\
{\mathcal V}(\nu,k,\tau)  &=& \sum_{\ell} (-i)^{\ell} (2\ell + 1) \, P_{\ell}(\nu)\, {\mathcal V}_{\ell}(k,\tau),  
\label{VV1}
\end{eqnarray}
where, following the same conventions of Eq. (\ref{r1}), ${\mathcal M}_{\ell}$, ${\mathcal N}_{\ell}$ and ${\mathcal V}_{\ell}$ denote the $\ell$-th multipole of the corresponding quantity. The result of the integration over $\nu$ is therefore
\begin{eqnarray}
&& \Sigma_{1}^{(\mathrm{t})}(k,\tau) = \frac{6}{35} {\mathcal N}_{4} - \frac{3}{7} {\mathcal N}_{2} - \frac{{\mathcal N}_{0}}{6} 
+ \frac{3}{10} \, i\, ( {\mathcal M}_{1} + {\mathcal M}_{3}),
\label{V16}\\
&& \Sigma_{2}^{(\mathrm{t})}(k,\tau) = - \frac{{\mathcal V}_{2}}{2} - \frac{{\mathcal V}_{4}}{4},
\label{V17}
\end{eqnarray}
The same considerations developed in the basis of the linear vector polarizations 
can be repeated in the case of the left and right polarized waves. Bearing in mind Eq. (\ref{BRDEC21}), Eqs. (\ref{V8})--(\ref{V11})
can be written:
\begin{eqnarray}
&& \Delta^{(\mathrm{v})}_{I}(\varphi,\mu,k,\tau) = 2 \sqrt{\frac{\pi}{3}} \biggl[ Y_{1}^{-1}(\mu,\varphi) {\mathcal M}_{L}(k,\tau) - Y_{1}^{-1}(\mu,\varphi) {\mathcal M}_{R}(k,\tau)\biggr],
\label{V18}\\
&& \Delta^{(\mathrm{v})}_{Q}(\varphi,\mu,k,\tau) =2 \mu \sqrt{\frac{\pi}{3}} \biggl[ Y_{1}^{-1}(\mu,\varphi,k,\tau) {\mathcal N}_{L}(k,\tau) - Y_{1}^{-1}(\mu,\varphi) {\mathcal N}_{R}(k,\tau)\biggr],
\label{V19}\\
&& \Delta^{(\mathrm{v})}_{U}(\varphi,\mu,k,\tau) = - 2 i \sqrt{\frac{\pi}{3}} \biggl[  Y_{1}^{-1}(\mu,\varphi) {\mathcal N}_{L}(k,\tau) + Y_{1}^{-1}(\mu,\varphi,k,\tau) 
{\mathcal N}_{R}(k,\tau) \biggr],
\label{V20}\\
&& \Delta^{(\mathrm{v})}_{V}(\varphi,\mu,k,\tau) = 2  \sqrt{\frac{\pi}{3}}\biggl[Y_{1}^{-1}(\mu,\varphi) {\mathcal V}_{L}(k,\tau) - Y_{1}^{-1}(\mu,\varphi) {\mathcal V}_{R}(k,\tau) \biggr].
\label{V21}
\end{eqnarray}
For sufficiently 
small angular scales (i.e. for sufficiently large multipoles) the microwave sky degenerates into a plane and the. In this situation 
microwave photons propagate, for all practical purposes, along the $\hat{z}$ axis and instead of the spherical decomposition 
based on spherical harmonics one can safely use a plane-wave decomposition. Since the wavelength of the photons 
is typically much shorter than the inhomogeneity scale of the magnetic field one could also argue, at this point, that the situation in which the magnetic field oriented along the direction of propagation of the relic (vector) wave is sufficiently generic. This conclusion should however be scrutinized more carefully and this is the purpose of the discussion reported hereunder.

Suppose that the direction of propagation of the vector wave is not parallel to the magnetic field direction but orthogonal. If the relic vector propagates along the magnetic field direction, then, in the language of 
Eq. (\ref{CS4}), $\hat{k} \parallel \hat{e}_{3}$ implying $\alpha=\beta =0$. If the relic vector propagates orthogonally 
to the magnetic field direction then we can set $\alpha = \beta = - \pi/2$ implying that $\hat{k} \perp \hat{e}_{3}$. 
The direction of $\hat{k}$ will still be chosen to be the $\hat{z}$ axis so that $(\hat{k} \cdot \hat{n}) = \mu = \cos{\vartheta}$.
In the case $\alpha = \beta = - \pi/2$ Eqs. (\ref{DIP4}), (\ref{DIP5}), (\ref{DIP6}) and (\ref{DIP7}) read
\begin{eqnarray}
M_{11}(\mu,\varphi,\nu,\varphi') &=& \zeta \Lambda_{1} \sqrt{1 - \mu^2} \sqrt{1 - \nu^2} + \zeta \Lambda_{1} \mu \nu \cos{\varphi'}  \cos{\varphi} - \Lambda_{3} \mu \nu \sin{\varphi} \sin{\varphi'} 
\nonumber\\
&+& i f_{\mathrm{e}}\Lambda_{2} \zeta ( \nu \sqrt{1 - \mu^2} \cos{\varphi'} - \mu \sqrt{ 1 - \nu^2} \cos{\varphi})
\label{V23}\\
 M_{12}(\mu,\varphi,\nu,\varphi') &=&  - \zeta \mu \Lambda_{1} \cos{\varphi} \sin{\varphi'}  - \Lambda_{3} \mu \cos{\varphi'} 
\sin{\varphi} 
\nonumber\\
&-& i f_{\mathrm{e}} \zeta \Lambda_{2} \sqrt{1 - \mu^2} \sin{\varphi'},
\label{V24}\\
M_{21}(\mu,\varphi,\nu,\varphi') &=&  - \Lambda_{3} \nu \cos{\varphi} \sin{\varphi'} - \zeta \Lambda_1 \nu \cos{\varphi'}\sin{\varphi} + i f_{\mathrm{e}} \zeta \Lambda_{2} \sqrt{1 - \nu^2},
\label{V25}\\
M_{22}(\mu,\varphi,\nu,\varphi') &=& \zeta \Lambda_{1}\sin{\varphi'} \sin{\varphi} - \Lambda_{3} \cos{\varphi'} \cos{\varphi}.
\label{V26}
\end{eqnarray}
In this case we can already expect, in comparison with the situation $\hat{k} \parallel \hat{e}_{3}$, that the  transport equations differ depending upon the 
specific vector polarization. The solution of the system can indeed be written as 
\begin{eqnarray}
&& \Delta^{(\mathrm{v})}_{I}(\varphi,\mu,k,\tau) = \sqrt{1 - \mu^2} \biggl[ \cos{\varphi} {\mathcal M}_{a}(k,\tau) + \sin{\varphi} {\mathcal M}_{b}(k,\tau)\biggr],
\label{V27}\\
&& \Delta^{(\mathrm{v})}_{Q}(\varphi,\mu,k,\tau) = \mu\sqrt{1 - \mu^2} \biggl[ \cos{\varphi} {\mathcal N}_{a}(k,\tau) + \sin{\varphi} {\mathcal N}_{b}(k,\tau)\biggr],
\label{V28}\\
&& \Delta^{(\mathrm{v})}_{U}(\varphi,\mu,k,\tau) = \sqrt{1 - \mu^2} \biggl[ -\sin{\varphi} {\mathcal N}_{a}(k,\tau) + \cos{\varphi} {\mathcal N}_{b}(k,\tau)\biggr],
\label{V29}\\
&& \Delta^{(\mathrm{v})}_{V}(\varphi,\mu,k,\tau) = \sqrt{1 - \mu^2}  
\sin{2\varphi} {\mathcal V}_{a}(k,\tau) + \mu {\mathcal V}_{b}(k,\tau).
\label{V30}
\end{eqnarray}
The azimuthal factorization of Eqs. (\ref{V27})--(\ref{V30}) 
is not arbitrary and it is dictated by the specific form of the system of 
Eqs. (\ref{V1})--(\ref{V4}) in the case when $\hat{k} \perp \hat{e}_{3}$.  
Starting with the polarization $W_{a}$ the corresponding evolution equations 
for ${\mathcal M}_{a}$, ${\mathcal N}_{a}$ and ${\mathcal V}_{a}$ are given by:
\begin{eqnarray}
&&\partial_{\tau} {\mathcal M}_{a} + ( i k \mu + \epsilon') {\mathcal M}_{a} 
+ i \mu \partial_{\tau} W_{a} - \epsilon' v_{a}  = 
- \epsilon' \mu \zeta^2 ( \Lambda_{1}^2 - f_{\mathrm{e}}^2 \Lambda_{2}^2) 
\Sigma^{(\mathrm{v})}_{1a},
\label{V31}\\
&& \partial_{\tau} {\mathcal N}_{a} + (i k \mu + \epsilon') {\mathcal N}_{a} 
= -\epsilon' \mu \zeta^2 ( \Lambda_{1}^2 - 
f_{\mathrm{e}}^2 \Lambda_{2}^2) \Sigma^{(\mathrm{v})}_{1a},
\label{V32}\\
&& \partial_{\tau} {\mathcal V}_{a} + ( i k \mu +\epsilon') {\mathcal V}_{a} = 0,
\label{V33}
\end{eqnarray}
where 
\begin{equation}
\Sigma^{(\mathrm{v})}_{1a} = \frac{3}{8} \int_{-1}^{1} d\nu[ \nu (\nu^2-1) {\mathcal M}_{a} + (\nu^4 -1) {\mathcal N}_{a}].
\label{V34}
\end{equation}
Consider then the case of the polarization $W_{b}$. The evolution equations 
are, in this second case, 
\begin{eqnarray}
&&\partial_{\tau} {\mathcal M}_{b} + ( i k \mu + \epsilon') {\mathcal M}_{b} 
+ i \mu \partial_{\tau} W_{b} - \epsilon' v_{b}  = 
 \epsilon' \mu \zeta \Lambda_{1} \Lambda_{3} \Sigma^{(\mathrm{v})}_{1b}
 \nonumber\\
&& - \frac{3}{4} f_{\mathrm{e}} \zeta \Lambda_{2} \Lambda_{3} \mu \int_{-1}^{1} \nu^2 
 {\mathcal V}_{b},
\label{V35}\\
&& \partial_{\tau} {\mathcal N}_{b} + (i k \mu + \epsilon') {\mathcal N}_{b} 
= 
\epsilon' \zeta \Lambda_{1} \Lambda_{3} \Sigma^{(\mathrm{v})}_{1b}
 \nonumber\\
&& - \frac{3}{4} f_{\mathrm{e}} \zeta \Lambda_{2} \Lambda_{3}  \int_{-1}^{1} \nu^2 
 {\mathcal V}_{b} d\nu,
\label{V36}\\
&& \partial_{\tau} {\mathcal V}_{b} + ( i k \mu +\epsilon') {\mathcal V}_{b} = 
\epsilon'  f_{\mathrm{e}} \mu \zeta \Lambda_{2} \Lambda_{3} \Sigma^{(\mathrm{v})}_{1b}
 \nonumber\\
&& - \frac{3}{4} f_{\mathrm{e}} \zeta \Lambda_{1} \Lambda_{3} \mu \int_{-1}^{1} \nu^2 
 {\mathcal V}_{b} d\nu,
\label{V37}
\end{eqnarray}
where 
\begin{equation}
\Sigma^{(\mathrm{v})}_{1b} = \frac{3}{8} \int_{-1}^{1} d\nu[ \nu (\nu^2-1) {\mathcal M}_{b} + (\nu^4 -1) {\mathcal N}_{b}].
\label{V38}
\end{equation}
The two sets of equations reported in Eqs. (\ref{V31})--(\ref{V33}) and in Eqs.  (\ref{V34})--(\ref{V37}) show various interesting features which can be summarized as 
follows:
\begin{itemize}
\item{} if $\hat{e}_{3} \parallel \hat{k}$ (i.e. $\alpha = \beta =0$) the evolution equations of the two vector polarizations are independent insofar as they can be given different initial conditions prior to decoupling but the evolution equations of the corresponding brightness perturbations are the same;
\item{} if $\hat{e}_{3} \perp \hat{k}$ (e.g. $\alpha = \beta = - \pi/2$) the two 
linear polarizations are equally independent but obey different evolution equations as it is clear by comparing Eqs. (\ref{V31})--(\ref{V33})
with Eqs.  (\ref{V35})--(\ref{V37});
\item{} the equation for the $a$-vector polarization (i.e. Eqs. (\ref{V31})--(\ref{V33})) lead to linear and circular photon polarizations which are a factor ${\mathcal O}(f_{\mathrm{e}})$ smaller than the corresponding equations for the $b$-vector polarization (see Eqs. (\ref{V34})--(\ref{V37})).
\end{itemize}

When $\hat{k} \perp \hat{e}_{3}$ and $\hat{k} = \hat{z}$ 
we also have that $\hat{a} = \hat{x}$ and $\hat{b}= \hat{y}$. But if 
$\alpha = \beta = - \pi/2$, then $\hat{e}_{3} = \hat{y}$. Therefore 
the amount of magnetically induced linear and circular photon polarization 
is larger when the magnetic field and the vector polarization are oriented 
along the same direction.
\begin{figure}[!ht]
\centering
\includegraphics[height=7cm]{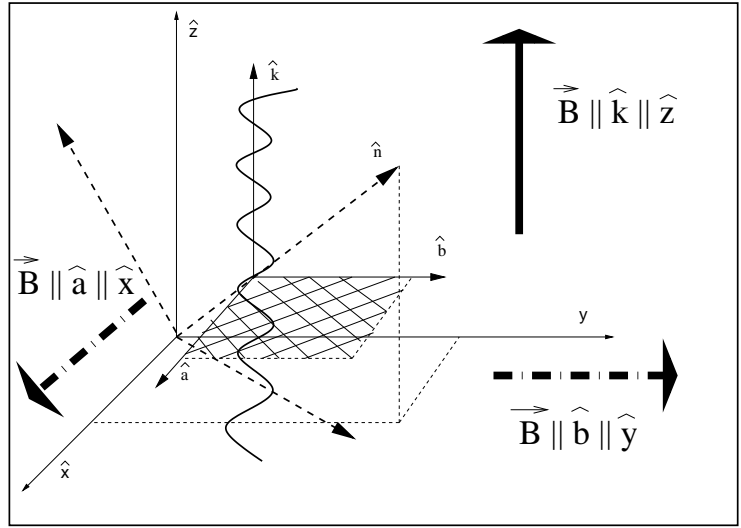}
\caption[a]{The interplay between the two linear vector polarizations (i.e. $\hat{a}$ and $\hat{b}$), the magnetic field direction and the direction of propagation of the scattered radiation (i.e. $\hat{n}$).}
\label{F2}      
\end{figure}
In Fig. \ref{F2} where the geometric set-up of the vector problem is summarized. 
The wiggly line represents pictorially a vector wave propagating in the direction 
$\hat{k}$ which has been taken to be aligned with the $\hat{z}$ axis. Always in Fig. 
\ref{F2} the shaded plane denotes the polarization plane of the vector wave 
spanned by the two unit vectors $\hat{a}$ and $\hat{b}$. Finally
$\hat{n}$ denotes the direction of propagation of the photons. 
If the direction of the magnetic field 
is parallel to the direction in which the vector modes propagate 
(thick arrow in Fig. \ref{F2}), the photons do not inherit a computable amount of circular polarization and, furthermore, the two linear vector polarizations 
will lead to the same transport equations for the brightness perturbations. 
Conversely, if 
the magnetic field is parallel to one of the two vector polarizations (thick dashed arrows in Fig. \ref{F2}) 
the transport equations for the two linear vector polarizations will be different.
If the linear vector polarization is aligned with the magnetic field 
intensity (for instance $\vec{B} \parallel \hat{b}$) the transport equations 
for ${\mathcal M}_{a}$, ${\mathcal N}_{a}$ and ${\mathcal V}_{a}$ will lead 
to a V-mode polarization larger than the one generated by the other vector polarization 
and described in terms of  ${\mathcal M}_{b}$, ${\mathcal N}_{b}$ and ${\mathcal V}_{b}$.

The source terms for the evolution equations of the vector modes can be averaged 
over the orientations of the magnetic field, i.e. 
\begin{eqnarray}
&& {\mathcal L}^{(\mathrm{v})}_{I}(\mu,\varphi,\vec{k},\tau) =  \epsilon' \hat{n}^{i} v^{(\mathrm{v})}_{i} 
\nonumber\\
&&+ \frac{3 \epsilon'}{128\pi^2} \int_{-1}^{1}d\nu \int_{0}^{2\pi}d\varphi' \int_{0}^{\pi} \sin{\beta}d\beta \int_{0}^{2\pi}d\alpha {\mathcal F}^{(\mathrm{v})}_{I}(\mu,\nu, \varphi,\varphi', \alpha,\beta),
\label{V1a}\\
&&  {\mathcal L}^{(\mathrm{v})}_{Q}(\mu,\varphi,\vec{k},\tau) = \frac{3 \epsilon'}{128 \pi^2}
\int_{-1}^{1}\,d\nu \int_{0}^{2 \pi}d\varphi'\int_{0}^{\pi} \sin{\beta} d \beta \int_{0}^{2\pi}d\alpha{\mathcal F}^{(\mathrm{v})}_{\mathrm{Q}}(\mu, \nu,\varphi, \varphi',\alpha,\beta),
\label{V2a}\\
&& {\mathcal L}^{(\mathrm{v})}_{U}(\mu,\varphi,\vec{k},\tau) =   \frac{3 \epsilon'}{128 \pi^2}
\int_{-1}^{1} d\nu \int_{0}^{2\pi}\, d\varphi'\, \int_{0}^{\pi} \sin{\beta} \, d \beta \int_{0}^{2\pi} d\alpha {\mathcal F}^{(\mathrm{v})}_{\mathrm{U}}(\mu,\nu,\varphi,\varphi',\alpha,\beta),
\label{V3a}\\
&& {\mathcal L}^{(\mathrm{v})}_{V}(\mu,\varphi,\vec{k},\tau) =  \frac{3 \epsilon'}{128 \pi^2}
\int_{-1}^{1} d\nu \int_{0}^{2\pi} d\varphi'\int_{0}^{\pi} \sin{\beta}d \beta \int_{0}^{2\pi}
 d\alpha {\mathcal F}^{(\mathrm{v})}_{\mathrm{V}}(\mu,\nu,\varphi,\varphi',\alpha,\beta).
\label{V4a}
\end{eqnarray}
The direct computation of the averaged source terms leads to the same 
expression for both vector polarizations. Denoting with ${\mathcal M}$ either 
${\mathcal M}_{a}$ or ${\mathcal M}_{b}$ (and similarly for ${\mathcal N}$ and 
${\mathcal V}$), Eqs. (\ref{V1a}), (\ref{V2a}), (\ref{V3a}) and (\ref{V4a}) lead to the 
following triplet of equations:
\begin{eqnarray}
\partial_{\tau}{\mathcal M} + (i k\mu + \epsilon') {\mathcal M} + i \mu \partial_{\tau} W - \epsilon' v &=& \frac{\mu}{15} \epsilon' \biggl[\biggl( 5 f_{\mathrm{e}}^2 \Lambda_{2}^2 - 7 \Lambda_{1}^2\biggr) 
\nonumber\\
 &+& 6 \zeta \Lambda_{1} \Lambda_{3} - 2 \Lambda_{3}^2 \biggr] \Sigma_{1}^{(\mathrm{v})},
\label{V5a}\\
  \partial_{\tau}{\mathcal N} + (i k\mu + \epsilon') {\mathcal N}  &=& \frac{\mu}{15} \epsilon' \biggl[\biggl( 5 f_{\mathrm{e}}^2 \Lambda_{2}^2 - 7 \Lambda_{1}^2\biggr)  
\nonumber\\ 
&+& 6 \zeta \Lambda_{1} \Lambda_{3} - 2 \Lambda_{3}^2 \biggr] \Sigma_{1}^{(\mathrm{v})},
\label{V6a}\\
\partial_{\tau} {\mathcal V} + ( i k \mu + \epsilon') {\mathcal V} &=& \frac{\epsilon'}{3} \zeta [ 2 \Lambda_{1} \Lambda_{3} - \zeta (\Lambda_{1}^2 + f_{\mathrm{e}}^2 \Lambda_{2}^2)] \Sigma_{2}^{(\mathrm{v})}.
\label{V7a}
\end{eqnarray}
As in the case of Eqs. (\ref{AVI})--(\ref{AVV}) 
if the magnetic field has a predominant direction over typical scales 
comparable with the wavelengths of the scattered photons, 
the circular polarization of the photons induced by the vector modes 
is larger than in the case where, over the same 
physical scales the magnetic field does not have a specific orientation. This conclusion can be 
reached by comparing Eqs. (\ref{V5a}), (\ref{V6a}) and (\ref{V7a}) to 
Eqs. (\ref{V34})--(\ref{V37}) obtained in the case when the magnetic field 
is oriented along one of the two polarizations of the relic vector.
\renewcommand{\theequation}{6.\arabic{equation}}
\setcounter{equation}{0}
\section{Tensor modes}
\label{sec6}
Recalling the notations introduced in Eqs. (\ref{BRDEC6a})--(\ref{BRDEC11b}), the evolution equations for the tensor components of the brightness perturbations shall be written, in general terms, as 
\begin{eqnarray}
&& {\mathcal L}^{(\mathrm{t})}_{I}(\mu,\varphi,\vec{k},\tau) = \frac{3 \epsilon'}{32\pi} \int_{-1}^{1} \,d\nu \int_{0}^{2\pi}\, d\varphi'\, {\mathcal F}^{(\mathrm{t})}_{I}(\mu,\nu, \varphi,\varphi', \alpha,\beta),
\label{T1}\\
&&  {\mathcal L}^{(\mathrm{t})}_{Q}(\mu,\varphi,\vec{k},\tau) = \frac{3 \epsilon'}{32 \pi}
\int_{-1}^{1}\,d\nu \int_{0}^{2 \pi}\, d\varphi'\,\,{\mathcal F}^{(\mathrm{t})}_{\mathrm{Q}}(\mu, \nu,\varphi, \varphi',\alpha,\beta),
\label{T2}\\
&& {\mathcal L}^{(\mathrm{t})}_{U}(\mu,\varphi,\vec{k},\tau) =   \frac{3 \epsilon'}{32 \pi}
\int_{-1}^{1} \,d\nu \int_{0}^{2\pi}\, d\varphi'\,\, {\mathcal F}^{(\mathrm{t})}_{\mathrm{U}}(\mu,\nu,\varphi,\varphi',\alpha,\beta),
\label{T3}\\
&& {\mathcal L}^{(\mathrm{t})}_{V}(\mu,\varphi,\vec{k},\tau) =  \frac{3 \epsilon'}{32 \pi}
\int_{-1}^{1} \,d\nu \int_{0}^{2\pi}\, d\varphi'\,\,{\mathcal F}^{(\mathrm{t})}_{\mathrm{V}}(\mu,\nu,\varphi,\varphi',\alpha,\beta).
\label{T4}
\end{eqnarray}
Consider first the case where the propagation of the relic graviton is parallel to the direction of the magnetic field intensity. 
The direction of propagation of the tensor wave can be chosen, without loss of generality, as $\hat{k} = \hat{z}$. Therefore
Eq. (\ref{CS4}) implies that $\alpha = \beta =0$, i.e. $\hat{k} \parallel \hat{e}_{3}$. 
To illustrate the azimuthal dependence of the problem it is instructive to write down Eq. (\ref{T1}) in explicit terms:
\begin{eqnarray}
&& \partial_{\tau} \Delta_{I}^{(\mathrm{t})} + ( i k \mu + \epsilon') \Delta_{I}^{(\mathrm{t})}  -\frac{1}{2}(1 - \mu^2) \biggl[ 
\cos{2\varphi} \partial_{\tau} h_{\oplus}+ \sin{2\varphi} \partial_{\tau} h_{\otimes}\biggr]
\nonumber\\
&& =   \frac{3 \epsilon'}{32\pi} \int_{-1}^{1} \,d\nu \int_{0}^{2\pi}\, d\varphi'\, {\mathcal F}^{(\mathrm{t})}_{I}(\mu,\nu, \varphi,\varphi', \alpha,\beta),
\label{T5}
\end{eqnarray}
where, according to Eq. (\ref{FI})  the integrand of the source term acts on the vector components of the various 
brightness perturbations and it is given by 
\begin{eqnarray}
&& {\mathcal F}^{(\mathrm{t})}_{I}(\mu,\nu, \varphi,\varphi', \alpha,\beta) = {\mathcal T}_{II}(\mu,\nu, \varphi,\varphi', \alpha,\beta) 
\Delta^{(\mathrm{t})}_{I}(\nu,\varphi') + 
 {\mathcal T}_{IQ}(\mu,\nu, \varphi,\varphi', \alpha,\beta)\Delta^{(\mathrm{t})}_{Q}(\nu,\varphi') 
\nonumber\\
&& +{\mathcal T}_{IU}(\mu,\nu, \varphi,\varphi', \alpha,\beta) \Delta^{(\mathrm{t})}_{U}(\nu,\varphi') 
+ {\mathcal T}_{IV}(\mu,\nu, \varphi,\varphi', \alpha,\beta) \Delta^{(\mathrm{t})}_{V}(\nu,\varphi').
\label{T6}
\end{eqnarray}
The remaining three equations (i.e. Eqs. (\ref{T2}), (\ref{T3}) and (\ref{T4})) have a similar structure 
but the contribution of the tensor modes of the geometry is absent. 
The azimuthal dependence can be decoupled from the radial 
dependence and the brightness perturbations will be 
\begin{eqnarray}
\Delta_{I}^{(\mathrm{t})}(\varphi,\mu,k,\tau) &=& (1 - \mu^2) \biggl[ \cos{2 \varphi} {\mathcal Z}_{\oplus}(\mu,k,\tau) + \sin{2 \varphi} {\mathcal Z}_{\otimes}(\mu, k, \tau)\biggr],
\label{T7}\\
\Delta_{Q}^{(\mathrm{t})}(\varphi,\mu,k,\tau) &=& (1 + \mu^2) \biggl[ \cos{2 \varphi} {\mathcal T}_{\oplus}(\mu,k,\tau) + \sin{2 \varphi} {\mathcal T}_{\otimes}(\mu,k,\tau)\biggr],
\label{T8}\\
\Delta_{U}^{(\mathrm{t})}(\varphi,\mu,k,\tau) &=& 2 \mu \biggl[ -\sin{2 \varphi} {\mathcal T}_{\oplus}(\mu,k,\tau) + \cos{2 \varphi} {\mathcal T}_{\otimes}(\mu,k,\tau)\biggr],
\label{T9}\\
\Delta_{V}^{(\mathrm{t})}(\varphi,\mu,k,\tau) &=& 2 \mu \biggl[ \cos{2 \varphi} {\mathcal S}_{\oplus}(\mu,k,\tau) + \sin{2 \varphi} {\mathcal S}_{\otimes}(\mu,k,\tau)\biggr].
\label{T10}
\end{eqnarray}
In the case $\hat{k} \parallel \hat{e}_{3}$ the symmetry of the system implies necessarily an ansatz in the form 
of Eqs. (\ref{T7}), (\ref{T8}), (\ref{T9}) and (\ref{T10}). Starting with the explicit form of Eq. (\ref{T5}), it is immediately 
clear that the form of $\Delta_{I}^{(\mathrm{t})}(\varphi,\mu,k,\tau)$ is constrained by the $\varphi$ dependence 
appearing at the left hand side of Eq. (\ref{T5}). The integrand at the right hand side of Eq. (\ref{T5}), i.e.
Eq. (\ref{T6}),  contains also $\Delta_{Q}^{(\mathrm{t})}(\varphi',\nu,k,\tau)$
whose explicit form is univocally determined by observing that the integral over $\varphi'$ must match with the $\varphi$ dependence 
appearing at the left hand side of Eq. (\ref{T5}). But the obtained ansatz for $\Delta_{I}^{(\mathrm{t})}(\varphi,\mu,k,\tau)$
and $\Delta_{Q}^{(\mathrm{t})}(\varphi,\mu,k,\tau)$ can be inserted back into Eq. (\ref{T2}): this step will constructively 
determine the explicit form of $\Delta_{U}^{(\mathrm{t})}(\varphi,\mu,k,\tau)$. Equation (\ref{T3}) will finally 
determine the explicit form of $\Delta_{V}^{(\mathrm{t})}(\varphi,\mu,k,\tau)$ whose $\varphi$ dependence 
will have to be consistent with Eq. (\ref{T4}). 
The result of this procedure, expressed by Eqs. (\ref{T7}), (\ref{T8}), (\ref{T9}) and (\ref{T10}), determines the radial evolution 
and, in particular,  the following set of equations \cite{mgpol4}:
\begin{eqnarray}
&&\partial_{\tau} {\mathcal Z} + (i k\mu + \epsilon') {\mathcal Z} - \frac{1}{2}\partial_{\tau} h   =  
\epsilon' \zeta^2(\omega) [\Lambda_{1}^2(\omega) - f_{\mathrm{e}}^2(\omega) \Lambda_{2}^2(\omega)] \Sigma^{(\mathrm{t})},
\label{T11}\\
&& \partial_{\tau}{\mathcal T} +  (i k\mu + \epsilon') {\mathcal T} + \epsilon' {\mathcal T} = -  \epsilon' \zeta^2(\omega) [\Lambda_{1}^2(\omega) - f_{\mathrm{e}}^2(\omega) \Lambda_{2}^2(\omega)] \Sigma^{(\mathrm{t})},
\label{T12}\\
&& \partial_{\tau}{\mathcal S} + (i k\mu + \epsilon') {\mathcal S} =0,
\label{T13}
\end{eqnarray}
where ${\mathcal Z}$, ${\mathcal T}$ and ${\mathcal S}$ denote either the $\oplus$ or the $\otimes$ polarization. 
By expanding ${\mathcal Z}$, ${\mathcal T}$  in series of Legendre polynomials 
 \begin{eqnarray}
&& {\mathcal Z}(\nu,k,\tau)  = \sum_{\ell} (-i)^{\ell} (2\ell + 1) \, P_{\ell}(\nu)\, {\mathcal Z}_{\ell}(k,\tau).
\label{ZZ1}\\
&& {\mathcal T}(\nu,k,\tau)  = \sum_{\ell} (-i)^{\ell} (2\ell + 1) \, P_{\ell}(\nu)\, {\mathcal T}_{\ell}(k,\tau),
\label{TT1}
\end{eqnarray}
the source term $\Sigma^{(\mathrm{t})}$ can also be expressed as
\begin{eqnarray}
\Sigma^{(\mathrm{t})} &=& \frac{3}{32} \int_{-1}^{1} d \nu 
[  (1 - \nu^2)^2 {\mathcal Z}(\nu)
- ( 1 + \nu^2)^2 {\mathcal T}(\nu) - 4 \nu^2 {\mathcal T}(\nu)] 
\nonumber\\
&=& \frac{3}{70}{\mathcal  Z}_{4} + \frac{{\mathcal Z}_{2}}{7} - \frac{{\mathcal Z}_{0}}{10}- \frac{3}{70} {\mathcal T}_{4} + \frac{6}{7} {\mathcal T}_{2} - \frac{3}{5} {\mathcal T}_{0},
\label{T14}
\end{eqnarray}
where, as usual, ${\mathcal Z}_{\ell}$ and ${\mathcal T}_{\ell}$ denote the $\ell$-th mulipoles of the corresponding functions. The results obtained in Eqs. (\ref{T11}), (\ref{T12}) and (\ref{T13}) with the partial treatment of the tensor modes developed in
\cite{mgpol4}. 
\begin{figure}[!ht]
\centering
\includegraphics[height=7cm]{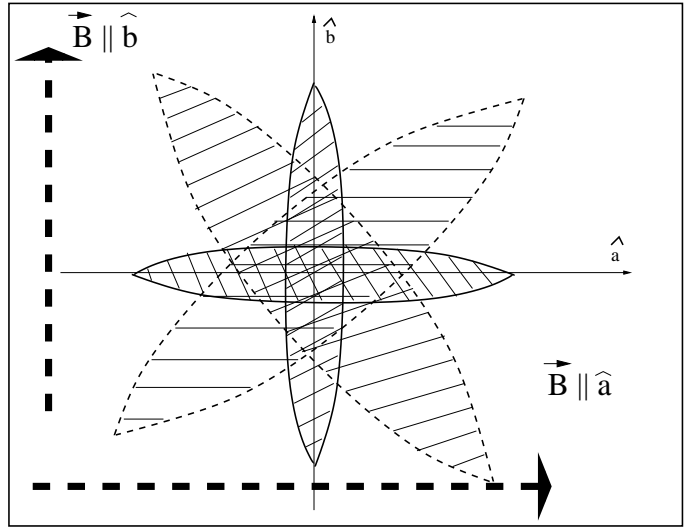}
\caption[a]{The $\oplus$ and $\otimes$ polarizations are illustrated, respectively, with full and dashed lines. The direction of propagation of the wave is not shown and it is orthogonal to the plane spanned by $\hat{a}$ and $\hat {b}$. When $\vec{B} \parallel \hat{k}$ the magnetic field is oriented perpendicularly to the plane of the figure.  }
\label{F3}      
\end{figure}
As in the case of the vectors instead of working with the linear polarizations of the relic gravitons, we could 
as well work with the circular polarization. The results obtained so far can be easily translated to the case when the relic gravitons are circularly polarized, always 
assuming that $\hat{k} \parallel \hat{e}_{3}$, i.e. that the direction of propagation 
of the relic gravitons is parallel to the orientation of the magnetic field intensity. 

 As expected from the vector case, when the relic tensor propagates orthogonally to the magnetic field direction, the two tensor polarizations will obey different equations but, at the same time, there will be differences in comparison 
with the vector case. The two polarizations 
of the relic gravitons when projected along the directions of the photon 
propagation will lead to a quadrupole term. The 
$\oplus$ and $\otimes$ polarization of the tensor mode are 
illustrated in Fig. \ref{F3} which should be compared 
with the shaded area of Fig. \ref{F2}.  When the $\oplus$ polarization propagates orthogonally to the magnetic field direction its evolution equations are given by:
\begin{eqnarray}
&& \partial_{\tau} {\mathcal Z}_{\oplus} + ( i k \mu + \epsilon') {\mathcal Z}_{\oplus} - \frac{1}{2} \partial_{\tau} h_{\oplus} = \frac{\epsilon'}{2} ( \zeta^2 \Lambda_{1}^2 + \Lambda_{3}^2) \Sigma^{(\mathrm{t})}_{\oplus}
\nonumber\\
&& + \frac{\epsilon'}{2(1- \mu^2) \cos{2 \varphi}} 
\biggl[ 
\Lambda_{3}^2 (1 + \mu^2) - \zeta^2 \biggl( \Lambda_{1}^2 ( 1 + \mu^2) - 2 
f_{\mathrm{e}}^2 \Lambda_{2}^2 ( \mu^2 -1)\biggr)\biggr] \Sigma^{(\mathrm{t})}_{\oplus},
\label{T15}\\
&& \partial_{\tau} {\mathcal T}_{\oplus} + ( i k \mu + \epsilon') {\mathcal T}_{\oplus} 
= - \frac{\epsilon'}{2} ( \zeta^2 \Lambda_{1}^2 + \Lambda_{3}^2) \Sigma^{(\mathrm{t})}_{\oplus}
\nonumber\\
&& - \frac{\epsilon'}{2(1- \mu^2) \cos{2 \varphi}} 
\biggl[ 
\Lambda_{3}^2 (1 + \mu^2) - \zeta^2 \biggl( \Lambda_{1}^2 ( 1 + \mu^2) - 2 
f_{\mathrm{e}}^2 \Lambda_{2}^2 ( \mu^2 -1)\biggr)\biggr] \Sigma^{(\mathrm{t})}_{\oplus},
\label{T16}\\
&& \partial_{\tau} {\mathcal S}_{\oplus} + ( i k \mu + \epsilon') {\mathcal S}_{\oplus} 
= - \frac{3}{16} f_{\mathrm{e}} \zeta^2 \Lambda_{1} \Lambda_{2} \frac{\sqrt{ 1 - \mu^2}}{\mu} \frac{\sin{\varphi}}{\cos{2 \varphi}} \Sigma^{(\mathrm{t})}_{\oplus},
\label{T17}
\end{eqnarray}
where it has been assumed that $\alpha = \beta = -\pi/2$.
When the $\otimes$ polarization propagates orthogonally to the magnetic field direction 
its evolution equations are given by:
\begin{eqnarray}
&& \partial_{\tau} {\mathcal Z}_{\otimes} + ( i k \mu + \epsilon') {\mathcal Z}_{\otimes} - \frac{1}{2} \partial_{\tau} h_{\otimes} = - \epsilon' \zeta \Lambda_{1} \Lambda_{3} \Sigma^{(\mathrm{t})}_{\otimes},
\label{T18}\\
&& \partial_{\tau} {\mathcal T}_{\otimes} + ( i k \mu + \epsilon') {\mathcal T}_{\otimes}  = - \epsilon' \zeta \Lambda_{1} \Lambda_{3} \Sigma^{(\mathrm{t})}_{\otimes},
\label{T19}\\
&& \partial_{\tau} {\mathcal S}_{\otimes} + ( i k \mu + \epsilon') {\mathcal S}_{\otimes}  = - \epsilon' f_{\mathrm{e}} \zeta \Lambda_{1} \Lambda_{3} \frac{\cos{\varphi}}{\sin{2 \varphi}} \frac{\sqrt{1 - \mu^2}}{\mu} \Sigma^{(\mathrm{t})}_{\otimes}.
\label{T20}
\end{eqnarray}
Equations (\ref{T15}), (\ref{T16}) and (\ref{T17}) can then  be compared 
to Eqs. (\ref{T18}), (\ref{T19}) and (\ref{T20}) recalling that, now the magnetic field 
is directed along $\hat{y}$.
The two tensor polarizations read $\hat{\epsilon}^{\oplus}_{ij} = (\hat{a}_{i} \hat{a}_{j} - \hat{b}_{i}\hat{b}_{j})$ 
and $\hat{\epsilon}^{\otimes}_{ij} = (\hat{a}_{i} \hat{b}_{j} + \hat{a}_{j} \hat{b}_{i})$. But 
since it has been assumed that $\hat{k} = \hat{z}$ we shall also have $\hat{a}=(1,\,0,\,0)= \hat{x}$ and $\hat{b}=(0,\,1,\,0)= \hat{y}$.
The addition of a magnetic field either along $\hat{a}$ or along $\hat{b}$ (i.e. orthogonally to $\hat{k}$)  is 
illustrated in Fig. \ref{F3} for the  two tensor polarizations. The polarization $\oplus$ spans the shaded area 
bounded by the full lines. The polarization $\otimes$ spans the shaded area bounded by the dashed 
lines. The effect of having an extra source of circular dichroism along the $\hat{a}$ (or along the $\hat{b}$ axis) 
will be felt by both tensor polarizations as quantitatively established in Eqs. (\ref{T15})--(\ref{T17}) and in Eqs. (\ref{T18})--(\ref{T20}).

The last step is to compute the evolution equations by averaging the source 
functions over the directions of the magnetic field. 
\begin{eqnarray}
&& {\mathcal L}^{(\mathrm{t})}_{I}(\mu,\varphi,\vec{k},\tau) = 
\frac{3 \epsilon'}{128\pi^2} \int_{-1}^{1}d\nu \int_{0}^{2\pi}d\varphi' \int_{0}^{\pi}
\sin{\beta}d\beta \int_{0}^{2\pi}d\alpha
{\mathcal F}^{(\mathrm{t})}_{I}(\mu,\nu, \varphi,\varphi', \alpha,\beta),
\label{T1a}\\
&&  {\mathcal L}^{(\mathrm{t})}_{Q}(\mu,\varphi,\vec{k},\tau) = \frac{3 \epsilon'}{128 \pi^2}
\int_{-1}^{1}d\nu \int_{0}^{2 \pi}d\varphi'\int_{0}^{\pi} \sin{\beta}d \beta \int_{0}^{2\pi}d\alpha
{\mathcal F}^{(\mathrm{t})}_{\mathrm{Q}}(\mu, \nu,\varphi, \varphi',\alpha,\beta),
\label{T2a}\\
&& {\mathcal L}^{(\mathrm{t})}_{U}(\mu,\varphi,\vec{k},\tau) =   \frac{3 \epsilon'}{128 \pi^2}
\int_{-1}^{1} d\nu \int_{0}^{2\pi}d\varphi'\int_{0}^{\pi} \sin{\beta}d \beta \int_{0}^{2\pi} d\alpha 
{\mathcal F}^{(\mathrm{t})}_{\mathrm{U}}(\mu,\nu,\varphi,\varphi',\alpha,\beta),
\label{T3a}\\
&& {\mathcal L}^{(\mathrm{t})}_{V}(\mu,\varphi,\vec{k},\tau) =  \frac{3 \epsilon'}{128 \pi^2}
\int_{-1}^{1} d\nu \int_{0}^{2\pi} d\varphi'\int_{0}^{\pi} \sin{\beta}d \beta \int_{0}^{2\pi}d\alpha
{\mathcal F}^{(\mathrm{t})}_{\mathrm{V}}(\mu,\nu,\varphi,\varphi',\alpha,\beta).
\label{T4a}
\end{eqnarray}
The result for the evolution equations of the tensor polarizations with averaged sources is given by:
\begin{eqnarray}
&& \partial_{\tau} {\mathcal Z} + ( i k \mu + \epsilon') {\mathcal Z} - \frac{1}{2} \partial_{\tau} h = \frac{\epsilon'}{15} [ \zeta^2 ( 7 \Lambda_{1}^2 - 5 f_{\mathrm{e}}^2 \Lambda_{2}^2) - 6 \zeta \Lambda_{1} \Lambda_{3} + 2 \Lambda_{3}^2] \Sigma^{(\mathrm{t})},
\label{T5a}\\
&& \partial_{\tau} {\mathcal T} + ( i k \mu + \epsilon') {\mathcal T} = - \frac{\epsilon'}{15} [ \zeta^2 ( 7 \Lambda_{1}^2 - 5 f_{\mathrm{e}}^2 \Lambda_{2}^2) - 6 \zeta \Lambda_{1} \Lambda_{3} + 2 \Lambda_{3}^2] \Sigma^{(\mathrm{t})},
\label{T6a}\\
&& \partial_{\tau} {\mathcal S} + ( i k \mu + \epsilon') {\mathcal S} =0
\label{T7a}
\end{eqnarray}
These results extend and partially correct the results derived in \cite{mgpol4}.
The correction has to do with the source term of Eq. (\ref{T7a}) which 
vanishes exactly unlike stated in \cite{mgpol4} because of an error 
in the azimuthal integrations. 

\renewcommand{\theequation}{7.\arabic{equation}}
\setcounter{equation}{0}
\section{Concluding remarks}
\label{sec7}
An arbitrarily oriented magnetic field has been incorporated  in the Stokes matrix of the last electron-photon scattering.
The transport equations for the scalar, vector and tensor components of the brightness perturbations have been 
derived and studied in various physical situations.  The obtained results pave the way for 
a consistent improvement of the available analytical and numerical tools used for the calculation of magnetized CMB anisotropies. The general treatment developed here is also expected to be relevant for the careful assessment of the level of circular polarization 
induced at last scattering by the presence of a small-scale component of the pre-decoupling magnetic field.

\newpage

\begin{appendix}
\renewcommand{\theequation}{A.\arabic{equation}}
\setcounter{equation}{0}
\section{Stokes and Mueller matrices}
\label{APPA}
The explicit form of the Stokes and Mueller matrices for arbitrary orientation of the magnetic field 
will now be reported. The four distinct entries of the Stokes matrix $M(\Omega,\Omega',\alpha,\beta)$ appearing in Eq. (\ref{EV1}) 
can be written, in explicit terms, as 
\begin{eqnarray}
&& M_{11}(\Omega,\Omega',\alpha,\beta) = \frac{\zeta \Lambda_{1} - \Lambda_{3}}{2} \biggl[ \sqrt{1 - \mu^2} \sqrt{1 - \nu^2} + \mu \nu \cos{(\varphi -\alpha)} \cos{(\varphi' - \alpha)}\biggr]
\nonumber\\
&&+ \frac{\zeta \Lambda_{1} + \Lambda_{3}}{2} \biggl\{\cos{2 \beta} \biggl[ \mu \nu \cos{(\varphi - \alpha)} \cos{(\varphi' - \alpha)} - \sqrt{ 1 - \mu^2} \sqrt{1 - \nu^2}\biggr]
\nonumber\\
&& +  \sin{2\beta} \biggl[ \mu 
\sqrt{1 - \nu^2} \cos{(\varphi - \alpha)} + \nu \sqrt{1 - \mu^2} \cos{(\varphi' -\alpha)}\biggr]\biggr\} 
\nonumber\\
&& + \zeta \Lambda_{1} \mu \nu \sin{(\varphi -\alpha)} \sin{(\varphi' - \alpha)} 
+ i f_{\mathrm{e}} \zeta \Lambda_{2} \biggl\{ \sin{\beta} \biggl[ \mu \sqrt{1 - \nu^2} \sin{(\varphi - \alpha)} 
\nonumber\\
&& - \nu \sqrt{1 - \mu^2} \sin{(\varphi' - \alpha)} \biggr] +
\mu \nu \cos{\beta} \sin{(\varphi - \varphi')}\biggr\},
\label{DIP4}\\
&& M_{12}(\Omega,\Omega',\alpha,\beta) = \frac{\Lambda_{3} - 
\Lambda_{1} \zeta}{2} \mu \sin{(\varphi' -\alpha)} \cos{(\varphi - \alpha)}
\nonumber\\ 
&&- \frac{\Lambda_{3} + \Lambda_{1} \zeta}{2} \biggl\{\mu \sin{(\varphi' -\alpha)} \cos{(\varphi - \alpha)} \cos{2\beta}
+ \sqrt{1 - \mu^2} \sin{(\varphi' - \alpha)} \sin{2\beta} \biggr\} 
\nonumber\\
&& + \zeta \Lambda_{1} \mu \sin{(\varphi - \alpha)} \cos{(\varphi' - \alpha)} 
\nonumber\\
&&- i f_{\mathrm{e}} \zeta \Lambda_{2} \biggl[ \mu \cos{\beta} \cos{(\varphi' - \varphi)}  + 
\sqrt{1 - \mu^2} \sin{\beta} \cos{(\varphi' - \alpha)}\biggr],
\label{DIP5}\\
&& M_{21}(\Omega,\Omega',\alpha,\beta)=  - \frac{\zeta \Lambda_{1} + \Lambda_{3}}{2} \sqrt{1 - \nu^2} \, \sin{2 \beta} \sin{(\varphi - \alpha)}
\nonumber\\
&&+ \frac{\nu}{4} (\zeta \Lambda_{1} + \Lambda_{3}) \biggl[ \sin{(\varphi + \varphi' - 2 \alpha)} - \sin{(\varphi' - \varphi)}\biggr]
\nonumber\\
&& - \frac{\nu}{4} (\zeta \Lambda_{1} + \Lambda_{3})\biggl[ \sin{(\varphi + \varphi' - 2 \alpha)} - \sin{(\varphi' - \varphi)}\biggr] \cos{2\beta}
\nonumber\\
&& i f_{\mathrm{e}} \Lambda_{2} \zeta \biggl[ \nu \cos{\beta} \cos{(\varphi' - \varphi)} + \sqrt{ 1 - \nu^2} \sin{\beta} \cos{(\varphi - \alpha)}\biggr]
\label{DIP6}\\
&& M_{22}(\Omega,\Omega',\alpha,\beta)= \frac{\zeta \Lambda_{1} + \Lambda_{3}}{4} \sin{2 \alpha} ( 1 - \cos{2 \beta}) \sin{(\varphi' + \varphi)}  - \Lambda_{3} \cos{(\varphi' - \varphi)}
\nonumber\\
&&+  \frac{\zeta \Lambda_{1} + \Lambda_{3}}{4} ( 1 + \cos{2\beta}) [ \cos{(\varphi' - \varphi)} - \cos{2 \alpha} \cos{(\varphi' + \varphi)}] 
\nonumber\\
&&+ \frac{\zeta \Lambda_{1} + \Lambda_{3}}{2}\biggl[ \cos{(\varphi' - \varphi)} + \cos{ 2 \alpha} \cos{(\varphi' + \varphi)}\biggr]
\nonumber\\
&& - i f_{\mathrm{e}} \zeta \Lambda_{2} \cos{\beta} \sin{(\varphi' - \varphi)}. 
\label{DIP7}
\end{eqnarray}
Equations (\ref{DIP4}), (\ref{DIP5}), (\ref{DIP6}) and (\ref{DIP7}) have various 
interesting limits which depend upon the specific orientation of the magnetic field 
intensity. In the absence of magnetic field we have that 
\begin{equation}
\Lambda_{1} \to 1, \qquad \Lambda_{2} \to 1,\qquad \Lambda_{3} \to 1,\qquad \zeta \to -1,
\qquad f_{\mathrm{e}} \to 0.
\label{LIM}
\end{equation}
In the limit defined by Eq. (\ref{LIM}), Eqs. (\ref{DIP4})--(\ref{DIP7}) reduce to
\begin{eqnarray}
&& M_{11}(\mu,\varphi,\nu,\varphi') = - \sqrt{ 1 - \mu^2} \sqrt{1 - \nu^2} - \mu \nu \cos{(\varphi'- \varphi)},
\nonumber\\
&& M_{12}(\mu,\varphi,\nu,\varphi') = \mu \sin{(\varphi' - \varphi)}, \qquad 
M_{21}(\mu,\varphi,\nu,\varphi')  = - \nu \sin{(\varphi' - \varphi)},
\nonumber\\
&& M_{22}(\mu,\varphi,\nu,\varphi')  = - \cos{(\varphi' - \varphi)},
\label{LIM2}
\end{eqnarray}
where, as already pointed out in section \ref{sec2}, $\mu = \cos{\vartheta}$ and $\nu = \cos{\vartheta'}$.
The integrands appearing in the source terms of 
Eqs. (\ref{EVS1}), (\ref{EVS2}), (\ref{EVS3} and (\ref{EVS4}) are given by 
\begin{eqnarray}
&&{\mathcal F}_{I}(\Omega, \Omega',\alpha,\beta)= {\mathcal T}_{II}(\Omega, \Omega',\alpha,\beta) + 
{\mathcal T}_{IQ}(\Omega, \Omega',\alpha,\beta) Q(\Omega') 
\nonumber\\
&& + {\mathcal T}_{IU}(\Omega, \Omega',\alpha,\beta) U(\Omega') + {\mathcal T}_{IV}(\Omega, \Omega',\alpha,\beta) V(\Omega'),
\label{FI}\\
&&{\mathcal F}_{Q}(\Omega, \Omega',\alpha,\beta) ={\mathcal T}_{QI}(\Omega, \Omega',\alpha,\beta)I(\Omega') + 
{\mathcal T}_{QQ}(\Omega, \Omega',\alpha,\beta)  Q(\Omega') 
\nonumber\\
&&+ {\mathcal T}_{QU}(\Omega, \Omega',\alpha,\beta) U(\Omega')  + 
{\mathcal T}_{QV}(\Omega, \Omega',\alpha,\beta) V(\Omega'),
\label{FQ}\\
&&{\mathcal F}_{U}(\Omega, \Omega',\alpha,\beta) ={\mathcal T}_{UI}(\Omega, \Omega',\alpha,\beta) I(\Omega') + 
{\mathcal T}_{UQ}(\Omega, \Omega',\alpha,\beta)Q(\Omega') 
\nonumber\\
&& + {\mathcal T}_{UU}(\Omega, \Omega',\alpha,\beta) U(\Omega') + 
{\mathcal T}_{UV}(\Omega, \Omega',\alpha,\beta) V(\Omega'),
\label{FU}\\
&& {\mathcal F}_{V}(\Omega, \Omega',\alpha,\beta)={\mathcal T}_{VI}(\Omega, \Omega',\alpha,\beta) I(\Omega') + 
{\mathcal T}_{VQ}(\Omega, \Omega',\alpha,\beta)  Q(\Omega') 
\nonumber\\
&&+ {\mathcal T}_{VU}(\Omega, \Omega',\alpha,\beta) U(\Omega') +  {\mathcal T}_{VV}(\Omega, \Omega',\alpha,\beta) V(\Omega'),
\label{FV}
\end{eqnarray}
where the matrix elements ${\mathcal T}_{ij}$ are computed in terms of Eqs. (\ref{DIP4}), (\ref{DIP5}), (\ref{DIP6}) and 
(\ref{DIP7}): 
\begin{eqnarray}
&& {\mathcal T}_{II}   = 2[\bigl| M_{11} \bigr|^2 + \bigl| M_{12} \bigr|^2 + \bigl| M_{21} \bigr|^2 + \bigl| M_{22} \bigr|^2],
\label{EV3}\\
&& {\mathcal T}_{IQ} =  2[\bigl| M_{11} \bigr|^2 - \bigl| M_{12} \bigr|^2 + \bigl| M_{21} \bigr|^2 - \bigl| M_{22} \bigr|^2], 
\label{EV4}\\
&& {\mathcal T}_{IU} =2[ M_{11} M_{12}^{*} + M_{12} M_{11}^{*} + M_{21} M_{22}^{*} + M_{22} M_{21}^{*}],
\label{EV5}\\
&& {\mathcal T}_{IV} = 2 [i \bigl( M_{12} M_{11}^{*}- M_{12}^{*} M_{11} + M_{22} M_{21}^{*} - M_{21} M_{22}^{*}],
\label{EV6}\\
&& {\mathcal T}_{QI} = 2[\bigl| M_{11} \bigr|^2 + \bigl| M_{12} \bigr|^2 - \bigl| M_{21} \bigr|^2 - \bigl| M_{22} \bigr|^2],
\label{EV7}\\
&& {\mathcal T}_{QQ} = 2[\bigl| M_{11} \bigr|^2 - \bigl| M_{12} \bigr|^2 - \bigl| M_{21} \bigr|^2 + \bigl| M_{22} \bigr|^2],
\label{EV8}\\
&& {\mathcal T}_{QU} = 2[M_{11} M_{12}^{*} + M_{12} M_{11}^{*} - M_{21} M_{22}^{*} - M_{22} M_{21}^{*}],
\label{EV9}\\
&& {\mathcal T}_{QV} =  2[i \bigl( M_{12} M_{11}^{*}- M_{12}^{*} M_{11} - M_{22} M_{21}^{*} + M_{21} M_{22}^{*}],
\label{EV10}\\
&& {\mathcal T}_{UI} = 2[M_{11} M_{21}^{*} + M_{12} M_{22}^{*} + M_{21} M_{11}^{*} + M_{22} M_{12}^{*}],
\label{EV11}\\
&& {\mathcal T}_{UQ} = 2[M_{11} M_{21}^{*} - M_{12} M_{22}^{*} + M_{21} M_{11}^{*} - M_{22} M_{12}^{*}], 
\label{EV12}\\
&& {\mathcal T}_{UU} = 2[M_{11} M_{22}^{*} + M_{12} M_{21}^{*}  + M_{21} M_{12}^{*} + M_{22} M_{11}^{*}],
\label{EV13}\\
&&  {\mathcal T}_{UV} = 2[i ( M_{12} M_{21}^{*} - M_{22}^{*} M_{11} + M_{22} M_{11}^{*} - M_{12}^{*} M_{21}],
\label{EV14}\\
&& {\mathcal T}_{VI} = 2[i \bigl( M_{11} M_{21}^{*} + M_{12} M_{22}^{*} - M_{21} M_{11}^{*} - M_{22} M_{12}^{*}],
\label{EV15}\\
&& {\mathcal T}_{VQ} = 2[i \bigl( M_{11} M_{21}^{*} - M_{12} M_{22}^{*} - M_{21} M_{11}^{*} + M_{22} M_{12}^{*} ],
\label{EV16}\\
&& {\mathcal T}_{VU}= 2[i \bigl( M_{11} M_{22}^{*} + M_{12} M_{21}^{*} - M_{21} M_{12}^{*} - M_{22} M_{11}^{*}],
\label{EV17}\\
&& {\mathcal T}_{VV} = 2[M_{22}^{*} M_{11} - M_{12} M_{21}^{*} + M_{22} M_{11}^{*} - M_{12}^{*} M_{21}].
\label{EV18}
\end{eqnarray}
In Eqs. (\ref{EV3})--(\ref{EV18}) the explicit dependence upon the six angles has been suppressed only 
for sake of simplicity. 
The components of the incident electric fields in the local frame $\hat{e}_{1}$, $\hat{e}_{2}$ and $\hat{e}_{3}$ 
can be related to the components off the electric field in the three Cartesian directions as 
\begin{eqnarray}
E_{1} &=& \cos{\alpha} \cos{\beta} E_{x}' + \sin{\alpha} \cos{\beta} E_{y}' - \sin{\beta} E_{z}',
\nonumber\\
E_{2} &=& -\sin{\alpha} E_{x}' +  \cos{\alpha} E_{y}' ,
\nonumber\\
E_{3} &=& \cos{\alpha} \sin{\beta} E_{x}' + \sin{\alpha} \sin{\beta} E_{y}' + \cos{\beta} E_{z}'.
\label{electric1}
\end{eqnarray}
The incident electric fields $E_{x}'$, $E_{y}'$ and $E_{z}'$ can be related, in turn, to their polar 
components as:
\begin{eqnarray}
E_{x}' &=& \cos{\vartheta'} \cos{\varphi'} E_{\vartheta}' - \sin{\varphi'} E_{\varphi}',
\nonumber\\
E_{y}' &=& \cos{\vartheta'} \sin{\varphi'} E_{\vartheta}' + \cos{\varphi'} E_{\varphi}',
\nonumber\\
E_{z}' &=& -\sin{\vartheta'} E_{\vartheta}',
\label{electric2}
\end{eqnarray}
where, as already spelled out in section \ref{sec2} the direction of propagation of 
the incident radiation $\hat{n}'$ coincides with $\hat{r}'$ and ($E_{\vartheta}'$,
$E_{\varphi}'$) are the components of the incident electric field in the spherical 
basis. The relations between the outgoing and the ingoing 
electric fields is given by 
\begin{eqnarray}
&& E_{\vartheta}(\vartheta,\varphi, \vartheta',\varphi',\alpha,\beta) = \frac{r_{\mathrm{e}}}{r} \biggl[ M_{11} E_{\vartheta}(\vartheta',\varphi') + 
M_{12} E_{\vartheta}(\vartheta',\varphi')  \biggr],
\nonumber\\
&& E_{\vartheta}(\vartheta,\varphi, \vartheta',\varphi',\alpha,\beta) = \frac{r_{\mathrm{e}}}{r} \biggl[ M_{21}E_{\vartheta}(\vartheta',\varphi') + 
M_{22} 
E_{\vartheta}(\vartheta',\varphi')  \biggr],
\label{electric}
\end{eqnarray}
where, $M_{ij} \equiv M_{ij}(\vartheta,\varphi, \vartheta',\varphi',\alpha,\beta)$ are given by Eqs. (\ref{DIP4})--(\ref{DIP6}).

\renewcommand{\theequation}{B.\arabic{equation}}
\setcounter{equation}{0}
\section{Angular integrations}
\label{APPB}
The angular integrations of the collision term can always be performed either directly or with the help of the appropriate 
Rayleigh expansion. The results are however rather lengthy. Furthermore, in the case of the vector and of the 
tensor modes (i.e. sections \ref{sec5} and \ref{sec6}), the integrations over $\varphi'$ change depending upon the 
the polarization of the vector (or of the tensor) wave. In the scalar case the results of the angular integration 
over $\varphi'$ are given hereunder following the notation of Eq. (\ref{SC1}):
\begin{eqnarray}
\overline{{\mathcal T}}_{I I} &=& 2 \pi \nu^2 \biggl\{ (3 \mu^2 -1) + f_{\mathrm{e}}^2 \biggl[ (1 + \mu^2) - 2 ( 1 + \mu^2) \sin^2{\beta} \cos^2{(\varphi - \alpha)} 
\nonumber\\
&+& \mu \sqrt{1 - \mu^2} \cos{(\varphi - \alpha)} \sin{2\beta}\biggr]\biggr\} + 
 2 \pi\biggl\{ 3 - \mu^2 + f_{\mathrm{e}}^2 \biggl[ ( 1 + \mu^2)  \nonumber\\
 &-& 2 
 \sin^2{\beta} ( 1 - \mu^2) \cos^2{(\varphi - \alpha)} 
 +\mu \sqrt{1 - \mu^2 } 
 \cos{(\varphi - \alpha)} \sin{2 \beta}\biggr]\biggr\}
 \label{TbarII}\\
 \overline{{\mathcal T}}_{I Q} &=& \frac{\pi}{2} ( \nu^2 -1) \biggl\{ 4( 3 \mu^2 -1) + 
 f_{\mathrm{e}}^2 \biggl[ 4 ( 1 - \mu^2) \sin^2{(\varphi- \alpha)} 
 \nonumber\\
 &+& (1 - \mu^2) \biggl( \cos{2(\varphi - \alpha-\beta)} + \cos{2 (\varphi - \alpha +\beta)}\biggr)
 + 2 \cos{2 \beta} ( 1 + 3 \mu^2) 
 \nonumber\\
 &-& 2 \mu \sqrt{1 - \mu^2}\biggl( \sin{ (\varphi - \alpha - 2\beta)} - \sin{(\varphi - \alpha - 2\beta)}\biggr)\biggr],
 \label{TbarIQ}\\
 \overline{{\mathcal T}}_{I U} &=&0,
 \label{TbarIU}\\
  \overline{{\mathcal T}}_{I V} &=& 8 \pi f_{\mathrm{e}} \nu [ ( 1 + \mu^2) \cos{\beta} 
  + \mu \sqrt{1 - \mu^2} \cos{(\varphi - \alpha)} \sin{\beta}],
  \label{TbarIV}\\
   \overline{{\mathcal T}}_{Q I} &=& 2 \pi \nu^2 \biggl\{ 3 (\mu^2 -1) + f_{\mathrm{e}}^2 \biggl[ (\mu^2 -1) +  2 \sin^2{\beta} ( 1 + (1 - \mu^2) \cos^2{(\varphi - \alpha)})
 \nonumber\\  
 &+&  \mu \sqrt{1 - \mu^2} \cos{(\varphi - \alpha)} \sin{2 \beta}\biggr]\biggr\} + 2 \pi \biggl\{ (1 - \mu^2) 
\nonumber\\
 &+& f_{\mathrm{e}}^2 \biggl[ (\mu^2 -1)+ 2 \sin^2{\beta} \biggl[ \sin^2{(\varphi- \alpha)} - \mu^2 \cos^2{(\varphi- \alpha)} \biggr] 
 \nonumber\\
&+& \mu \sqrt{1 - \mu^2} \cos{(\varphi - \alpha)} \sin{2\beta}\biggl\}
  \label{TbarQI}\\
  \overline{{\mathcal T}}_{Q Q} &=& \frac{\pi}{2} ( 1 -\nu^2) \biggl\{ 12 (1 -\mu^2) +
    f_{\mathrm{e}}^2 \biggl[ 2 (\mu^2 -1) 
 \nonumber\\
 &-&2 ( 1 + \mu^2) \cos{2(\varphi -\alpha)} + (1 + \mu^2) ( \cos{2(\varphi - \alpha -\beta})+ \cos{2(\varphi -\alpha +\beta)})
 \nonumber\\
 &+&   6 \cos{2\beta}( 1 - \mu^2) + 2 \mu \sqrt{1 - \mu^2} (\sin{(\varphi -\alpha - 2\beta)} - \sin{(\varphi -\alpha + 2\beta)})\biggr] \biggr\}.
 \label{TbarQQ}\\
  \overline{{\mathcal T}}_{Q U} &=&0,
 \label{TbarQU}\\
   \overline{{\mathcal T}}_{Q V} &=& 8\pi f_{\mathrm{e}} \nu [ (\mu^2 -1) \cos{\beta} + \mu \sqrt{1 -\mu^2} \cos{(\varphi -\alpha)} \sin{\beta}],
 \label{TbarQV}\\
 \overline{{\mathcal T}}_{U I} &=& - 2\pi f_{\mathrm{e}}^2 \sin{(\varphi -\alpha)} \biggl\{ 
 \nu^2 \biggl[ 4 \mu \cos{(\varphi -\alpha)} \sin^2{\beta} + 
 \sqrt{1 - \mu^2} \sin{2 \beta}\biggr]
 \nonumber\\
 &+& \biggl[ \sqrt{1 - \mu^2 } \sin{2\beta} - 4 \mu \cos{(\varphi - \alpha)} \sin^2{\beta}\biggr]\biggr\}
 \label{TbarUI}\\
 \overline{{\mathcal T}}_{U Q} &=& - 4 \pi f_{\mathrm{e}}^2 (\nu^2 - 1) \sin{\beta} 
 \sin{(\varphi - \alpha)} \biggl[ \sqrt{1- \mu^2} \cos{\beta} + 2 \mu \cos{(\varphi -\alpha)}
 \sin{\beta}\biggr],
 \label{TbarUQ}\\
   \overline{{\mathcal T}}_{U U} &=&0,
 \label{TbarUU}\\
  \overline{{\mathcal T}}_{U V} &=& - 8 \pi \nu
   f_{\mathrm{e}} \sqrt{ 1 - \mu^2} \sin{(\varphi -\alpha)} \sin{\beta}
  \label{TbarUV}\\
  \overline{{\mathcal T}}_{V I} &=& 4 \pi f_{\mathrm{e}}
  \biggl\{ \nu^2 \biggl[ 2 \mu \cos{\beta} - \sqrt{1-\mu^2} 
  \cos{(\varphi -\alpha)} \sin{\beta} \biggr]
 \nonumber\\ 
 &+& \biggl[ 2 \mu \cos{\beta} + 3 \sqrt{1 - \mu^2} \cos{(\varphi -\alpha)} 
 \sin{\beta} \biggr]\biggr\},
\label{TbarVI}\\
\overline{{\mathcal T}}_{V Q} &=& 2 \pi f_{\mathrm{e}} (\nu^2 -1) \biggl[ 4 \mu \cos{\beta} -2 \sqrt{1 - \mu^2} \cos{(\varphi -\alpha)} \sin{\beta}\biggr],
\label{TbarVQ}\\
\overline{{\mathcal T}}_{V U} &=& 0,
\label{TbarVU}\\
\overline{{\mathcal T}}_{V V} &=& 8\pi \nu \biggl\{ \mu +\cos{\beta} f_{\mathrm{e}}^2 
\biggl[ \mu \cos{\beta} + \sqrt{1 - \mu^2 } \cos{(\varphi -\alpha)} \sin{\beta}\biggr] 
\biggr\}.
\label{TbarVV}
 \end{eqnarray}
\end{appendix}

\end{document}